\documentclass[12pt,a4paper]{article}

\newcommand{\ppbar}{p^{\!\!\!\!\!\textsuperscript{\tiny{(--)}}}\!\!}
\newcommand{\ptmiss}{{p\!\!\!\!\! \not \,\,\,\,}_T}
\newcommand{\mumu}{\mu^\pm \mu^\pm}
\newcommand{\ee}{e^\pm e^\pm}
\newcommand{\mue}{\mu^\pm e^\pm}

\usepackage{amsmath,amssymb,graphicx}
\usepackage{epsfig}
\usepackage{cite}

\begin{document}

\begin{center}
\begin{Large}
{\bf Heavy neutrino signals at large hadron colliders}

\end{Large}

\vspace{0.5cm}
F. del Aguila$^1$, J. A. Aguilar--Saavedra$^1$, R. Pittau$^2$  \\[0.2cm] 
{\it $^1$ Departamento de F\'{\i}sica Te\'orica y del Cosmos and CAFPE, \\
Universidad de Granada, E-18071 Granada, Spain} \\[0.1cm]
{\it $^2$ Dipartimento di Fisica Teorica, Universit\`a di Torino, 
and INFN \\ Sezione di Torino, V. Pietro Giuria 1, I-10125 Torino, Italy} 
\end{center}

\begin{abstract}
We study the LHC discovery potential for heavy Majorana neutrino singlets in
the process $pp \to W^+ \to \ell^+ N \to \ell^+ \ell^+ jj$ ($\ell=e,\mu$)
plus its charge conjugate. With a fast detector simulation we show that
backgrounds involving two like-sign charged
leptons are not negligible and, moreover, they cannot be eliminated with simple
sequential
kinematical cuts. Using a likelihood analysis it is shown that, for heavy
neutrinos coupling only to the muon, LHC has $5\sigma$ sensitivity for masses
up to 200 GeV in the final state $\mu^\pm \mu^\pm jj$. This reduction in
sensitivity, compared to previous parton-level estimates, is driven by the
$\sim 10^2-10^3$ times larger background. Limits are also provided for
$e^\pm e^\pm jj$ and $e^\pm \mu^\pm jj$ final states, as well as for Tevatron.
For heavy Dirac neutrinos the prospects are worse because backgrounds involving
two opposite charge leptons are much larger. For this case, we study the
observability of the lepton flavour violating signal $e^\pm \mu^\mp jj$. As a
by-product of our analysis, heavy neutrino production has been implemented
within the ALPGEN framework.
\end{abstract}

\section{Introduction}

Large hadron colliders involve strong interacting particles as initial states,
giving rise to huge hadronic cross sections. The large luminosities
expected will also provide quite large electroweak signals, with for instance 
$1.6 \times 10^{10}$ ($4 \times 10^7$) $W$ bosons at LHC (Tevatron) for a
luminosity of 100 (2) fb$^{-1}$. Therefore, these colliders can be used for
precise studies of the leptonic sector as well, and
in particular they can produce new
heavy neutrinos at an observable level, or improve present limits on their
masses and mixings
\cite{Datta:1993nm,Almeida:2000pz,Panella:2001wq,Han:2006ip} (see
Ref.~\cite{delAguila:2006dx} for a review). 
These new fermions transform trivially under the gauge symmetry group 
of the Standard Model (SM), and in the absence of other interactions 
they are produced and decay only through their mixing with the SM leptons. 
With new
interactions, like for instance in left-right models \cite{Langacker:1984dc},
heavy neutrinos can be produced by gauge couplings 
unsuppressed by small mixing angles, yielding larger cross sections and  
implying a much higher collider discovery reach
\cite{Ferrari:2000sp,Gninenko:2006br,Keung:1983uu,Datta:1992qw}.
Heavy neutrinos could also be copiously
produced in pairs through the exchange of a relatively light
$Z'$ boson~\cite{delAguila:2007ua}.
In these scenarios, however, the observation of the new interactions could be
more interesting than the existence of new heavy neutrinos. 

We will concentrate on the first possibility and neglect other new production
mechanims, taking a conservative approach. In this case, for example,
it has been claimed by looking at the lepton number violating (LNV)
$\Delta L = 2$ process $p\ppbar \rightarrow \mu^\pm \mu^\pm jj$ that 
LHC will be sensitive to heavy Majorana neutrinos with masses $m_N$
up to 400 GeV, whereas Tevatron is sensitive to masses up to 150 GeV
\cite{Almeida:2000pz,Han:2006ip}. 
However, as we will show, taking into account the actual backgrounds these
limits are far from being realistic.
In particular, backgrounds involving $b$ quarks,
including for instance $t \bar t n j$ (with $nj$
standing for $n=0,1,2,\dots$ additional jets), are two orders of magnitude
larger than previously estimated.
Moreover, in the region $m_N < M_W$ the largest and irreducible background is
$b \bar b nj$, by far dominant but overlooked in previous parton-level analyses
\cite{Han:2006ip}.
 In this work we make a detailed study, at the level of
fast simulation, of the LHC sensitivity to Majorana neutrinos in the process 
$pp \rightarrow \mu^\pm\mu^\pm jj$, which is the cleanest final state,
for both $m_N > M_W$ and $m_N < M_W$.
We also study the processes $pp \to e^\pm e^\pm jj$ and
$pp \to e^\pm \mu^\pm jj$ for which the sensitivity is slightly worse. Heavy
Dirac neutrinos do not produce LNV signals and then their observation is much
more difficult. As an example, we examine the lepton flavour violating (LFV)
signal $e^\pm \mu^\mp jj$, produced by a heavy Dirac neutrino coupling to the
electron and muon.

The generation of heavy neutrino signals has been implemented in the
ALPGEN \cite{Mangano:2002ea} framework, including the process studied here
as well as other final states. In the following, after making precise our
assumptions and notation in section~\ref{sec:2}, we describe the
implementation of heavy neutrino production in ALPGEN in section~\ref{sec:3}.
We present our detailed results in section~\ref{sec:4}, where we will eventually
find that heavy neutrinos can be discovered up to masses of the order of
200 GeV, and that for $N$ lighter than the $W$ boson its mixing can be probed
at the $10^{-2}$ level (for a ``reference'' mass $m_N = 60$ GeV).
These figures are much less optimistic than
in previous literature. Estimates for
Tevatron are given in section~\ref{sec:5}, and our conclusions are
drawn in section~\ref{sec:6}. In two appendices we detail the evaluation of the
$b \bar b nj$ background and the heavy neutrino mass reconstruction,
respectively.

\section{Heavy neutrino interactions}
\label{sec:2}

Our assumptions and notation are reviewed in more detail in
Ref.~\cite{delAguila:2006dx}
(see also  Refs.~\cite{Mohapatra:1998rq,Branco:1999fs}).
The SM is only extended with heavy neutrino singlets $N_j$, which
are assumed to have masses of the order of the electroweak scale,
up to few hundreds of GeV.
We concentrate on the lightest one, assuming for
simplicity that the other extra neutrinos are heavy enough to
neglect possible interference effects. The new heavy neutrino
$N$ (where we suppress the unnecessary subindex)
can have Dirac character, what requires the addition of at least two
singlets, or Majorana, in which case
$(N_L)^c \equiv C N_L^T = N_R$ and lepton number is violated.
In either case it is produced and decays
through its mixing with the light leptons, which is described
by the interaction Lagrangian (in standard notation)
\begin{eqnarray}
\mathcal{L}_W & = &
- \frac{g}{\sqrt 2}  \left( \bar \ell \gamma^\mu V_{\ell N}
P_L N \; W_\mu + \bar N \gamma^\mu V_{\ell N}^* P_L \ell \; W_\mu^\dagger
\right) \,, \nonumber \\
\mathcal{L}_Z & = &
- \frac{g}{2 c_W}  \left( \bar \nu_\ell \gamma^\mu
V_{\ell N} P_L N + \bar N \gamma^\mu V_{\ell N}^* P_L \nu_\ell \right)
Z_\mu \,, \nonumber \\
\mathcal{L}_H & = &
- \frac{g \, m_N}{2 M_W} \, \left( \bar \nu_\ell \, V_{\ell N}
P_R N + \bar N \, V_{\ell N}^* P_L \nu_\ell \right) H \,.
\label{ec:nNZ}
\end{eqnarray}
The SM Lagrangian remains unchanged in the limit of small
mixing angles $V_{\ell N}$, $\ell = e, \mu, \tau$ (which is the
actual case), up to very small corrections $O(V^2)$. 
Neutral couplings involving two heavy neutrinos are also of order $V^2$.
The heavy neutrino mass $m_N$ joins two different bispinors
in the Dirac case and the same one in the Majorana case.
Heavy neutrino decays are given by their interactions in Eqs.~(\ref{ec:nNZ}):
$N \to W^+ \ell^-$, $N \to Z \nu$, $N \to H \nu$,
plus $N \to W^- \ell^+$ for a heavy Majorana neutrino.
For $m_N < M_W$ all these decays produce three body final
states, mediated by off-shell $W$, $Z$ or $H$ bosons.
The total width for a Majorana neutrino is twice larger than for a Dirac
one with the same couplings
\cite{Gluza:1996bz,Pilaftsis:1991ug,delAguila:2005pf,delAguila:2006dx}.

As it is apparent from Eqs.~(\ref{ec:nNZ}), heavy neutrino signals are
proportional to the neutrino mixing with the SM leptons $V_{\ell N}$.
Limits on these matrix elements have been extensively discussed
in previous literature, and we quote here only the main results. Low-energy data
constrain the quantities
\begin{equation}
\Omega_{\ell \ell'} \equiv \delta_{\ell \ell'} - \sum_{i=1}^3 V_{\ell \nu_i}
V_{\ell' \nu_i}^* = \sum_{j=1}^n V_{\ell N_j} V_{\ell' N_j}^* \,.
\label{bounds1}
\end{equation}
A global fit to tree level processes
involving light neutrinos as external states gives
\cite{Bergmann:1998rg,Bekman:2002zk},
\begin{equation}
\Omega_{ee} \leq 0.0054 \,, \quad \Omega_{\mu \mu} \leq 0.0096 \,, \quad
\Omega_{\tau \tau} \leq 0.016
\label{eps1}
\end{equation}
at 90\% confidence level (CL). Note that a global fit
without the unitarity bounds implies $\Omega_{ee} \leq 0.012$
\cite{Bergmann:1998rg}.
Additionally, for Majorana neutrinos coupling to the electron the experimental
bound on neutrinoless double beta decay requires
\cite{Aalseth:2004hb}
\begin{equation}
\left| \, \sum_{j=1}^n \, V_{e N_j}^2 \, \frac{1}{m_{N_j}} \, \right|
\, <  5 \times 10^{-8} \; {\rm GeV}^{-1} \,.
\label{beta}
\end{equation}%
If $V_{e N_j}$ saturate $\Omega_{ee}$ in Eq. (\ref{eps1}),
this limit can be satisfied either demanding that $m_{N_j}$ are large enough,
beyond the TeV scale \cite{Benes:2005hn}
and then beyond LHC reach, or that there is a
cancellation among the different terms in Eq.~(\ref{beta}), as
may happen in definite models \cite{Ingelman:1993ve},
in particular for (quasi)Dirac
neutrinos.

Flavour changing neutral processes further restrict $\Omega_{\ell \ell'}$.
The new contributions, and then the bounds, depend on the heavy neutrino
masses.
In the limit $m_{N_j}^2 \gg M_W^2 \gg |V_{\ell N_j}|^2 m_{N_j}^2$
\footnote{When $V_{\ell N_j} > M_W/m_{N_j}$ the non-decoupling terms in the amplitude, proportional to $V^4_{\ell N_j} m^2_{N_j}/M_W^2$, cannot be neglected
because they dominate over the $V_{\ell N_j}^2$ terms \cite{Ilakovac:1994kj}.}
they imply \cite{Tommasini:1995ii} 
\begin{equation}
|\Omega_{e \mu}| \leq 0.0001 \,, \quad |\Omega_{e \tau}| \leq 0.01 \,, \quad
|\Omega_{\mu \tau}| \leq 0.01 \,.
\label{eps2}
\end{equation}
Except in the case of $\Omega_{e \mu}$, for which experimental constraints
on lepton flavour violation are rather stringent, these limits are similar
to the limits on the diagonal elements. An important difference, however, is
that (partial) cancellations among loop contributions of different
heavy neutrinos may be at work \cite{delAguila:2005mf}.
Cancellations with other new physics contributions are also possible. Since
we are interested in determining the heavy neutrino discovery potential and
the direct limits on neutrino masses and mixings which can be eventually
established, we must consider the largest possible neutrino mixings, although
they may require model dependent cancellations or fine-tuning.

\section{Heavy neutrino production with ALPGEN}
\label{sec:3}

For the signal event generation we have extended
ALPGEN~\cite{Mangano:2002ea} with heavy neutrino production. This Monte Carlo
generator evaluates tree level SM processes and provides unweighted events
suitable for simulation. A simple way of including new processes
taking advantage of the ALPGEN framework is to provide the corresponding
squared amplitudes decomposed as a sum over the different colour structures. 
In the case of heavy neutrinos this is trivial because there is only 
one term. This method requires to evaluate from the beginning the squared
amplitudes for the processes one is interested in, what is done using
HELAS~\cite{helas}. An alternative
possibility which gives more flexibility for future applications is to
implement the new vertices at the same level as the SM ones, what is quite more
involved.

\begin{figure}[htb]
\begin{center}
\begin{tabular}{ccc}
\epsfig{file=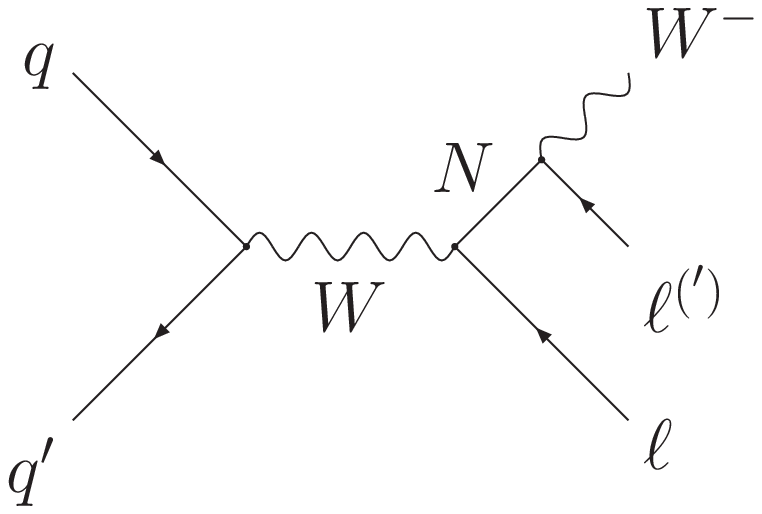,height=3cm,clip=} & \hspace{1cm} &
\epsfig{file=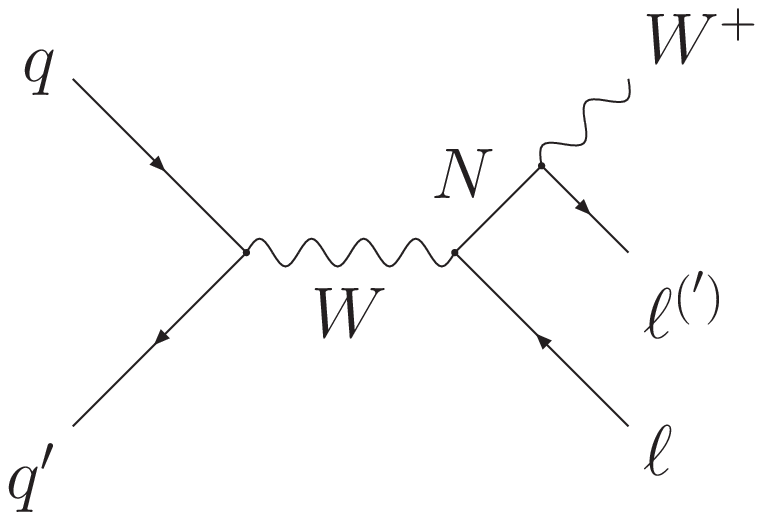,height=3cm,clip=} \\
(a) & & (b) \\
\end{tabular}
\caption{Feynman diagrams for the process $q \bar q' \to \ell^+ N$, followed by
LNV decay $N \to \ell^{(')+ }W^-$ (a) and lepton number conserving (LNC)
decay $N \to \ell^{(')-} W^+$ (b).
The diagrams for the charge conjugate processes are similar.}
\label{fig:diag}
\end{center}
\end{figure}

We have restricted ourselves to single heavy neutrino production.
Pair production is suppressed by an extra $V^2$ mixing factor and by 
the larger center of mass energy required, what implies smaller
PDFs and more suppressed $s$-channel propagators.
Single heavy neutrino production can proceed through $s$-channel $W, Z$ or $H$ 
exchange. The first two production mechanisms have been implemented in ALPGEN
for the various possible final states given by the heavy neutrino decays
$N \to W^\pm \ell^\mp$, $N \to Z \nu_\ell$, $N \to H \nu_\ell$ with
$\ell=e,\mu,\tau$, and for both Dirac or Majorana $N$.
In the case $m_N < M_W$ all decays
are three-body, and mediated by off-shell $W$, $Z$ or $H$. The transition from
two-body to three-body decays on the $M_W$, $M_Z$ and $M_H$ thresholds is
smooth, since the calculation of matrix elements and the $N$ width are done for
off-shell intermediate bosons. Two approximations are made, however.
The small mixing of heavy neutrinos with charged leptons implies that
their production is dominated
by diagrams with $N$ on-shell, like those shown in Fig.~\ref{fig:diag}, with a
pole enhancement factor, and that non-resonant diagrams are negligible.
(Additionally, to isolate heavy neutrino signals from the background one
expects that the heavy neutrino mass will have to be reconstructed to
some extent.) Then, the only diagrams included are the resonant ones.
In the calculation we also neglect light fermion masses except for the bottom
quark.

\begin{figure}[htb]
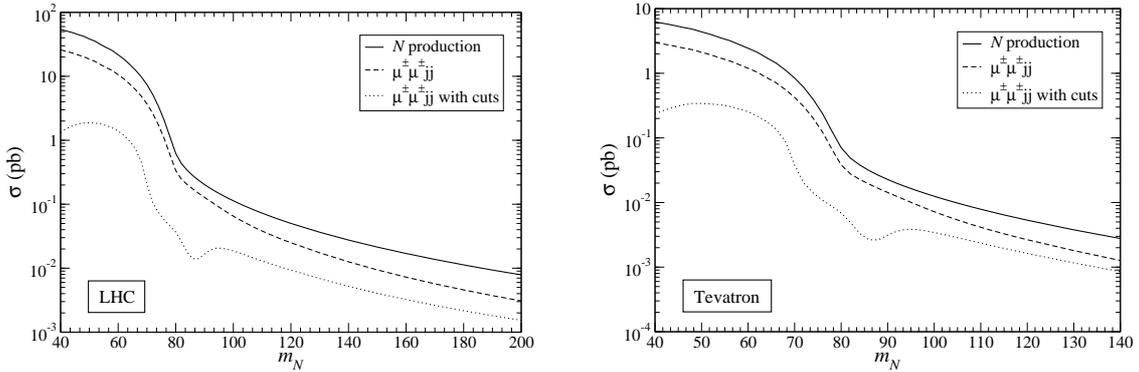

\begin{center}
\begin{tabular}{ccc}
\epsfig{file=Figs/cross-LHC.eps,height=4.9cm,clip=} & &
\epsfig{file=Figs/cross-Tev.eps,height=4.9cm,clip=} 
\end{tabular}
\caption{Cross sections for heavy neutrino production at LHC (left) and Tevatron
(right), as a function of the heavy neutrino mass, for $|V_{\mu N}| = 0.098$.
The solid lines correspond to total $\mu N$ cross section, the dashed lines
include the decay to like-sign muons and the dotted lines are the same but
including the kinematical cuts in Eq.~(\ref{ec:gcuts}).}
\label{fig:cross}
\end{center}
\end{figure}

Generator-level results are presented in Fig.~\ref{fig:cross} for LHC and
Tevatron in the relevant mass ranges. Solid lines correspond to the total
$\mu N$ cross sections for $|V_{\mu N}| = 0.098$, $V_{eN}=V_{\tau N} = 0$.
The dashed lines are the cross sections for the final state
$\mu^\pm \mu^\pm jj$,
which is the cleanest one. The  dotted lines are the same but with 
kinematical cuts 
\begin{align}
\mathrm{LHC}: \begin{array}{ccc} p_T^\mu \geq 10~\mathrm{GeV}\,, &
 |\eta^\mu| \leq 2.5\,, & \Delta R_{\mu j} \geq 0.4\,,  \\
p_T^j \geq 10~\mathrm{GeV}\,, & |\eta^j| \leq 2.5\,,
\end{array} \notag \\
\mathrm{Tevatron}: \begin{array}{ccc} p_T^\mu \geq 10~\mathrm{GeV}\,, &
 |\eta^\mu| \leq 2\,, & \Delta R_{\mu j} \geq 0.4\,,  \\
p_T^j \geq 10~\mathrm{GeV}\,, & |\eta^j| \leq 2.5\,,
\end{array}
\label{ec:gcuts}
\end{align}
included to reproduce roughly the
acceptance of the detector and give approximately the ``effective'' size of the
observable
signal. Of course, the correct procedure is to perform a simulation, as we do
in next section, but for illustrative purposes we include the
cross-sections after cuts. In particular, they clearly show that
although for $m_N < M_W$ the total cross sections grow several orders of
magnitude, both at LHC and Tevatron, partons tend to be produced with low
transverse momenta (the two muons and two
quarks result from the decay of an on-shell $W$), making the observable signal
much smaller. These results are in agreement with those previously obtained in
Ref.~\cite{Han:2006ip}.

\section{Di-lepton signals at LHC}
\label{sec:4}

The most interesting scenario for LHC is when the heavy neutrino has Majorana
nature and couples only to the muon, so that it produces a
final state $\mu^\pm \mu^\pm jj$ with two same sign muons and at least two jets.
Since this LNV signal has sometimes been considered \cite{Almeida:2000pz,
Han:2006ip} to be almost background free (more realistic background estimates
are given in Ref.~\cite{Dreiner:2000vf}), a detailed discussion of the actual
backgrounds is worthwhile. A first group of processes
involves the production of additional leptons,
either neutrinos or charged leptons (which may be missed in the detector). 
The main ones are $W^\pm W^\pm nj$ and $W^\pm Z nj$.
We point out that not only the processes
with $n=2$ contribute:
processes with $n<2$ are backgrounds due to the appearance of extra jets
from pile-up, and processes with $n>2$ cannot be cleanly removed because of
pile-up on the signal. A second group includes final states with $b$ and/or
$\bar b$ quarks, like $t\bar t nj$, with
semileptonic decay of the $t\bar t$ pair, and $Wb \bar b nj$, with $W$
decaying leptonically. In these cases the additional like-sign
muon results from the decay of a $b$ or $\bar b$ quark. 
Only a tiny fraction of
such decays produce isolated muons with sufficiently high transverse momentum.
But, since the $t \bar tnj$ and $W b \bar b nj$ cross sections are so large,
these backgrounds are also much larger than 
backgrounds with two weak gauge bosons.
Finally, $b \bar b nj$ production is several orders of magnitude
larger than all processes mentioned above,
but the produced muons have small $p_T$ and invariant mass in this
case. Then, in general it might be eliminated with suitable high-$p_T$ cuts on
charged leptons \cite{Abulencia:2007rd} (see section~\ref{sec:150}), but for
$m_N < M_W$ the heavy neutrino signal is also characterised by very small
transverse momenta (see section~\ref{sec:60}), and this background turns out
to be the dominant one. The same applies for $c \bar c nj$, but with the
difference that $c$ quark decays produce isolated charged leptons much less
often than $b$ decays.

Other LNV signals produced by heavy neutrinos are $e^\pm e^\pm jj$
and $e^\pm \mu^\pm jj$. They have the same SM backgrounds but with one
important difference: $b$ decays produce ``apparently isolated'' electrons more
often than muons, because electrons are detected in the calorimeter while
muons travel to the muon chamber. Hence, the corresponding backgrounds
$t \bar t nj, b \bar b nj \to e^\pm e^\pm X/e^\pm \mu^\pm X$ are larger than
the ones involving only muons. A precise evaluation of these backgrounds,
optimising the criteria for electron isolation, seems to require a full
simulation of the detector. 
The limits provided in these cases must be regarded with some caution in this
respect, and should be confirmed with a full detector simulation.

We have generated the signal and backgrounds using ALPGEN and
passing them through PYTHIA 6.4 \cite{pythia} with the MLM prescription
\cite{mlm} to avoid double counting of jet radiation. A fast simulation of the
ATLAS detector \cite{atlfast} has been performed.
For the signal and all
backgrounds except $b \bar b nj$ and $c \bar c nj$ the number of simulated
events corresponds to at least
10 times the luminosity considered (which is 30 fb$^{-1}$), so as to reduce
statistical fluctuations, and the number of events is scaled accordingly. For
$b \bar b nj$ and $c \bar c nj$ the luminosity simulated is 0.075 fb$^{-1}$.
Their evaluation is further discussed in appendix~\ref{sec:a}.
It must also be noted that in the signal simulation all $W$ decays in
$p p \to \ell N \to \ell \ell' W$ are included. Leptonic $W$ decays give an
extra $\sim 20\%$ contribution to di-lepton final states when the
charged lepton from the $W$ decay is missed, or when $W$ decays to
$\tau \nu$ and the tau lepton decays hadronically.

\subsection{$\ell^\pm \ell^\pm jj$ production for $m_N > M_W$}
\label{sec:150}

In this mass region we take the reference values $m_N = 150$ GeV and
(a) $V_{\mu N} = 0.098$, $V_{eN} = V_{\tau N} = 0$; (b) $V_{e N} = 0.073$,
$V_{\mu N} = V_{\tau N} = 0$; (c) $V_{e N} = 0.073$, $V_{\mu N} = 0.098$,
$V_{\tau N} = 0$.
The pre-selection criteria used for our analysis are: 
\begin{itemize}
\item[(i)] two like-sign isolated charged leptons with 
pseudorapidity $|\eta^\ell| \leq 2.5$ and transverse momentum $p_T^\ell$
larger than 10 GeV (muons) or 15 GeV (electrons), and no additional isolated
charged leptons;
\item[(ii)] no additional non-isolated muons; 
\item[(iii)] two jets with $|\eta^j| \leq 2.5$ and $p_T^j \geq 20$ GeV. 
\end{itemize}
We point out that for $\mumu jj$ final states the requirement (ii)
reduces the backgrounds involving $Z$ bosons by almost a factor of two,
and thus proves to be quite useful. The number 
of events at LHC for 30 fb$^{-1}$ after pre-selection cuts is given in 
Table~\ref{tab:Nsb150}. 
Additional backgrounds such as $t \bar b$, $t \bar t t \bar t$,
$t \bar t b \bar b$, $Z t \bar t nj$,
$WWZnj$, $WZZnj$ and $ZZZnj$ are smaller and we do not show them, 
but they are included in the estimation of the signal significance below.
The number of like-sign dimuon events from $c \bar c nj$ displayed between
parentheses corresponds to an estimation, because no $\mumu X$ events are found
in the sample simulated (more details can be found in
appendix~\ref{sec:a}).
We also note that the higher $p_T$ threshold for electrons contributes to the
difference between the numbers of $\ee jj$ and $\mue jj$ events, which are
expected to be similar in some cases, for example for $t \bar t nj$.

\begin{table}[htb]
\begin{center}
\begin{tabular}{cccccccc}
& \multicolumn{3}{c}{Pre-selection} & \hspace{.5cm} &
  \multicolumn{3}{c}{Selection} \\
                 & $\mumu$  & $\ee$    & $\mue$ 
	     &   & $\mumu$  & $\ee$    & $\mue$ \\
$N~(\mathrm{a})$ & 113.6    & 0        & 0 
             &   & 59.1     & 0        & 0  \\
$N~(\mathrm{b})$ & 0        & 72.0     & 0 
             &   & 0        & 17.6     & 0  \\
$N~(\mathrm{c})$ & 78.4     & 25.5     & 82.6 
             &   & 41.6     & 4.7      & 22.4 \\
$b \bar b nj$    & 14800    & 52000    & 82000 
             &   & 0        & 0        & 0   \\
$c \bar c nj$    & (11)     & 300      & 200 
             &   & (0)        & 0        & 0   \\
$t \bar t nj$    & 1162.1   & 8133.0   & 15625.3
             &   & 2.4      & 8.3      & 7.7 \\
$tj$             & 60.8     & 176.5    & 461.5
             &   & 0.0      & 0.0      & 0.1 \\
$W b \bar b nj$  & 124.9    & 346.7    & 927.3
             &   & 0.4      & 0.6      & 0.3 \\
$W t \bar t nj$  & 75.7     & 87.2     & 166.9
             &   & 0.3      & 0.0      & 0.0 \\
$Z b \bar b nj$  & 12.2     & 68.9     & 117.0
             &   & 0.0      & 0.2      & 0.0 \\
$WW nj$          & 82.8     & 89.0     & 174.8
             &   & 0.5      & 0.1      & 0.7 \\
$WZ nj$          & 162.4    & 252.0    & 409.2
             &   & 4.8      & 1.8      & 2.3 \\
$ZZ nj$          & 3.8      & 13.3     & 12.9
             &   & 0.0      & 0.6      & 0.1 \\
$WWW nj$         & 31.9     & 30.1     & 64.8
             &   & 0.9      & 0.1      & 0.0    
\end{tabular}
\caption{Number of $\ell^\pm \ell^\pm jj$ events at LHC for 30 fb$^{-1}$, at the
pre-selection and selection levels. The heavy
neutrino signal is evaluated assuming $m_N = 150$ GeV and coupling (a) to the
muon, $V_{\mu N} = 0.098$; (b) to the electron, $V_{e N} = 0.073$; (c) to both,
$V_{e N} = 0.073$ and $V_{\mu N} = 0.098$.}
\label{tab:Nsb150}
\end{center}
\end{table}

\begin{figure}[htb]
\begin{center}
\epsfig{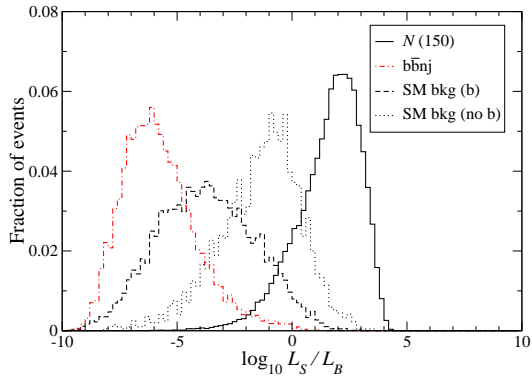} \\
\caption{Log-likelihood function for the
$\mumu jj$ signal with $m_N = 150$ GeV and its backgrounds.}
\label{fig:logLS150}
\end{center}
\end{figure}

Let us concentrate on $\mumu jj$ final states.
The fast simulation shows that SM backgrounds are about two orders of
magnitude larger than previously estimated (three orders if we include
$b \bar b nj$). Moreover, they cannot be sufficiently 
suppressed with respect to the heavy neutrino signal using simple cuts.
Some obvious discriminating variables are:
\begin{itemize}
\item The missing momentum $\ptmiss$. It is smaller for the signal because
it does not have neutrinos in the final state, but nonzero due to energy
mismeasurement in the detector.
\item The separation between the muon with smallest $p_T$ (we label the two
muons as $\mu_1$, $\mu_2$, by decreasing transverse momentum) and the closest
jet, $\Delta R_{\mu_2 j}$. 
For backgrounds involving high-$p_T$ $b$ quarks this separation tends to be
rather small.
\item The transverse momentum of the two muons, $p_T^{\mu_1}$ and 
$p_T^{\mu_2}$, respectively.
In particular $p_T^{\mu_2}$ 
is a good discriminant against backgrounds from $b$ quarks, because
these typically have one muon  with small $p_T$.
\end{itemize}
\begin{figure}[htb]
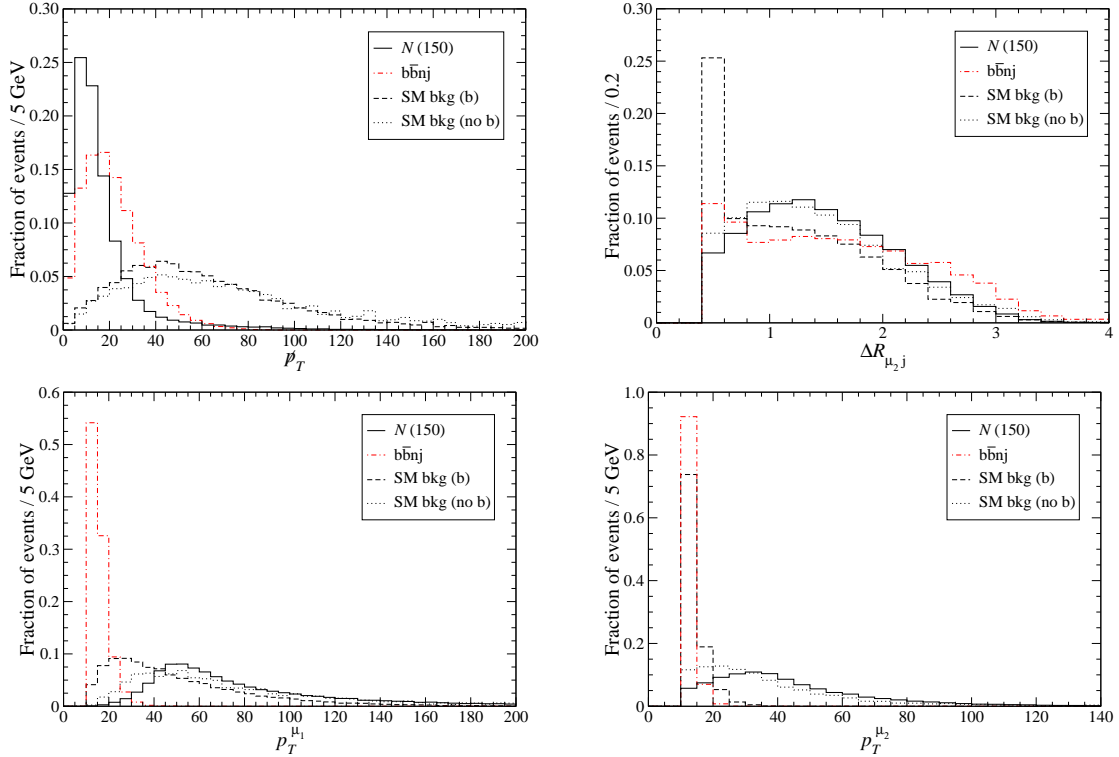

\begin{center}
\begin{tabular}{ccc}
\epsfig{file=Figs/ptmiss-150.eps,height=4.9cm,clip=} & &
\epsfig{file=Figs/dRl2j-150.eps,height=4.9cm,clip=} \\
\epsfig{file=Figs/ptl1-150.eps,height=4.9cm,clip=} & &
\epsfig{file=Figs/ptl2-150.eps,height=4.9cm,clip=}
\end{tabular}
\caption{Normalised distributions of several discriminating variables for the
$\mumu jj$ signal with $m_N = 150$ GeV and its backgrounds (see the text).}
\label{fig:var1}
\end{center}
\end{figure}
\begin{figure}[p]
\begin{center}
\begin{tabular}{ccc}
\epsfig{file=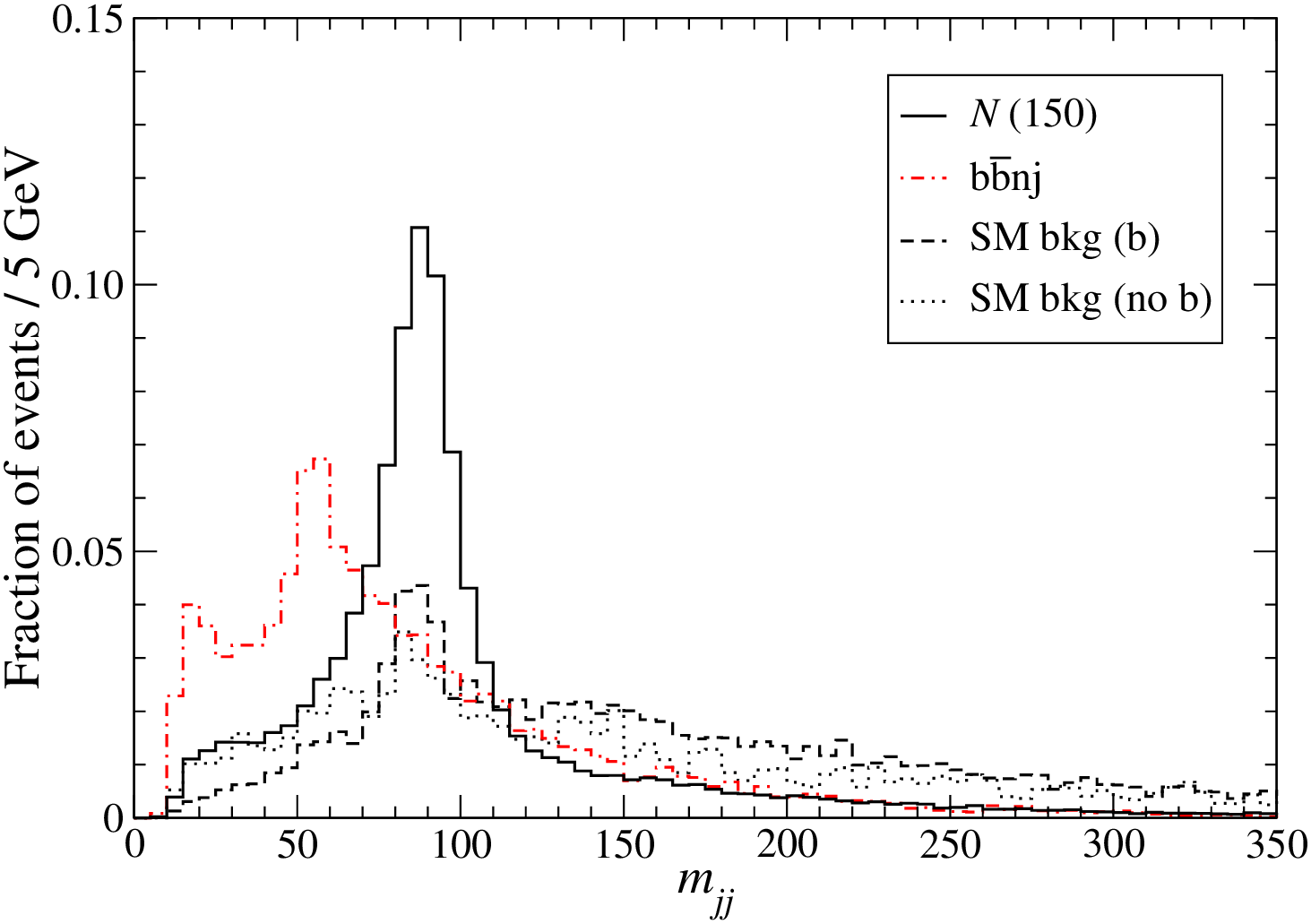,height=4.9cm,clip=} & &
\epsfig{file=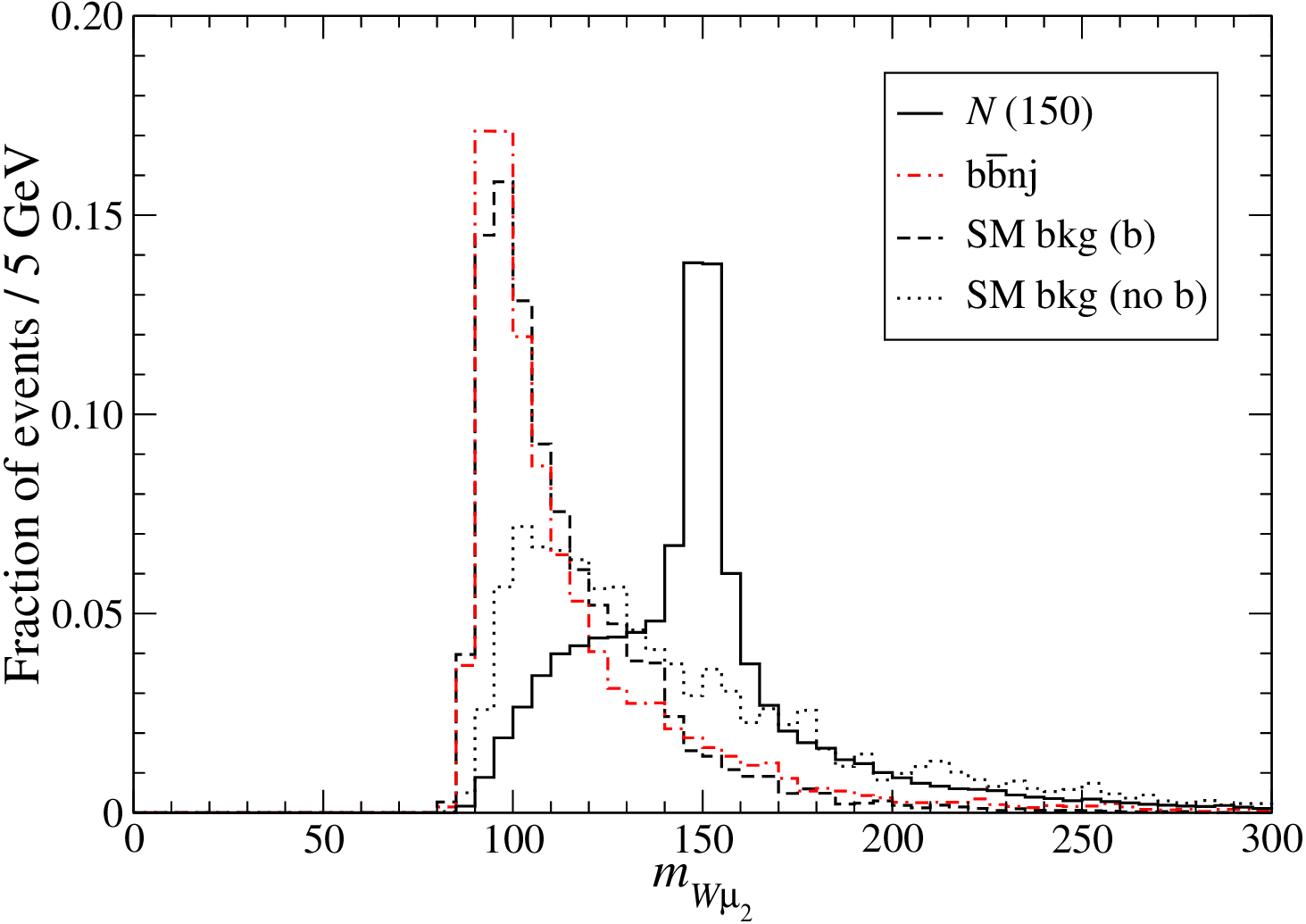,height=4.9cm,clip=} \\
\epsfig{file=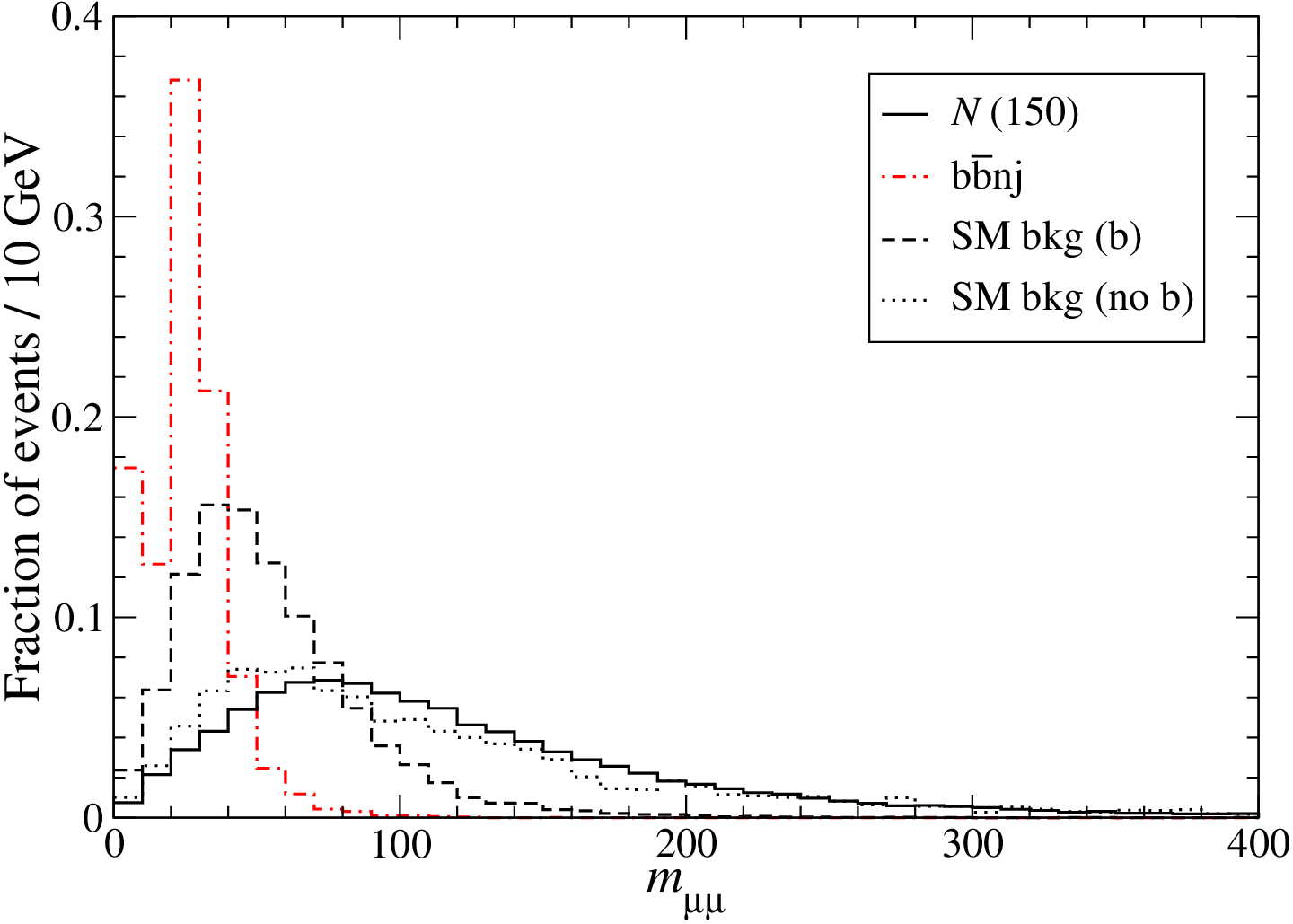,height=4.9cm,clip=} & &
\epsfig{file=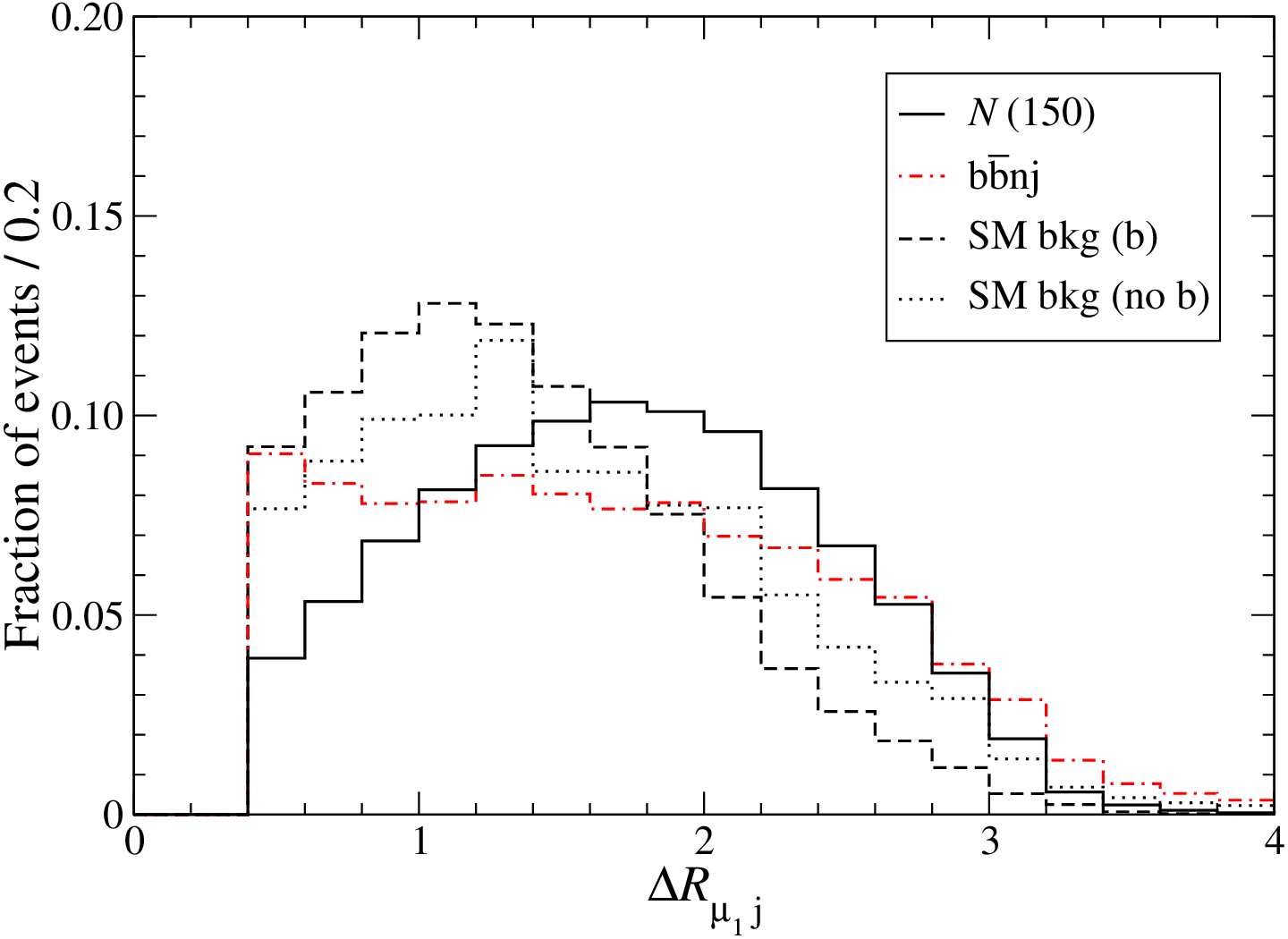,height=4.9cm,clip=} \\
\epsfig{file=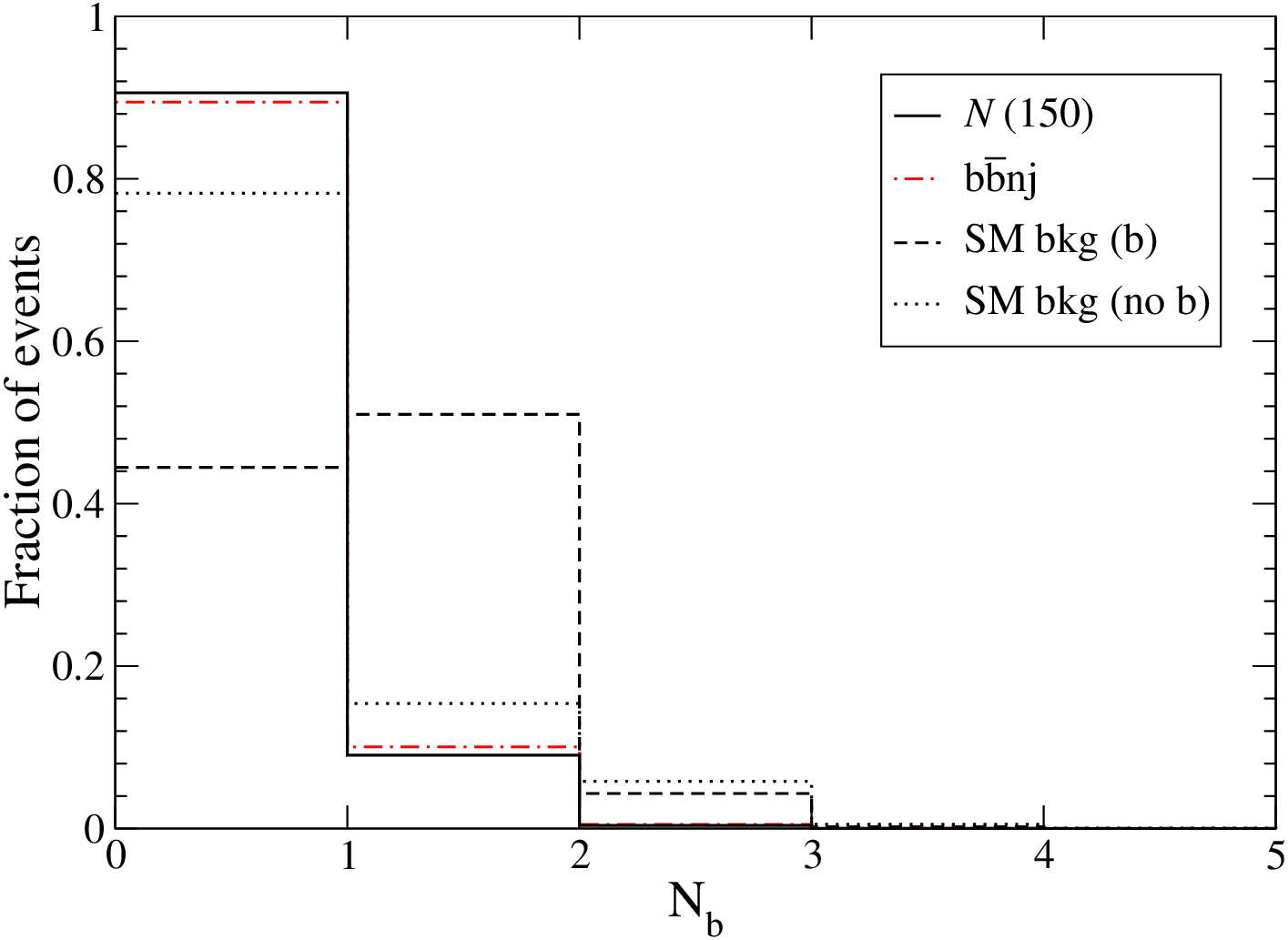,height=4.9cm,clip=} & & 
\epsfig{file=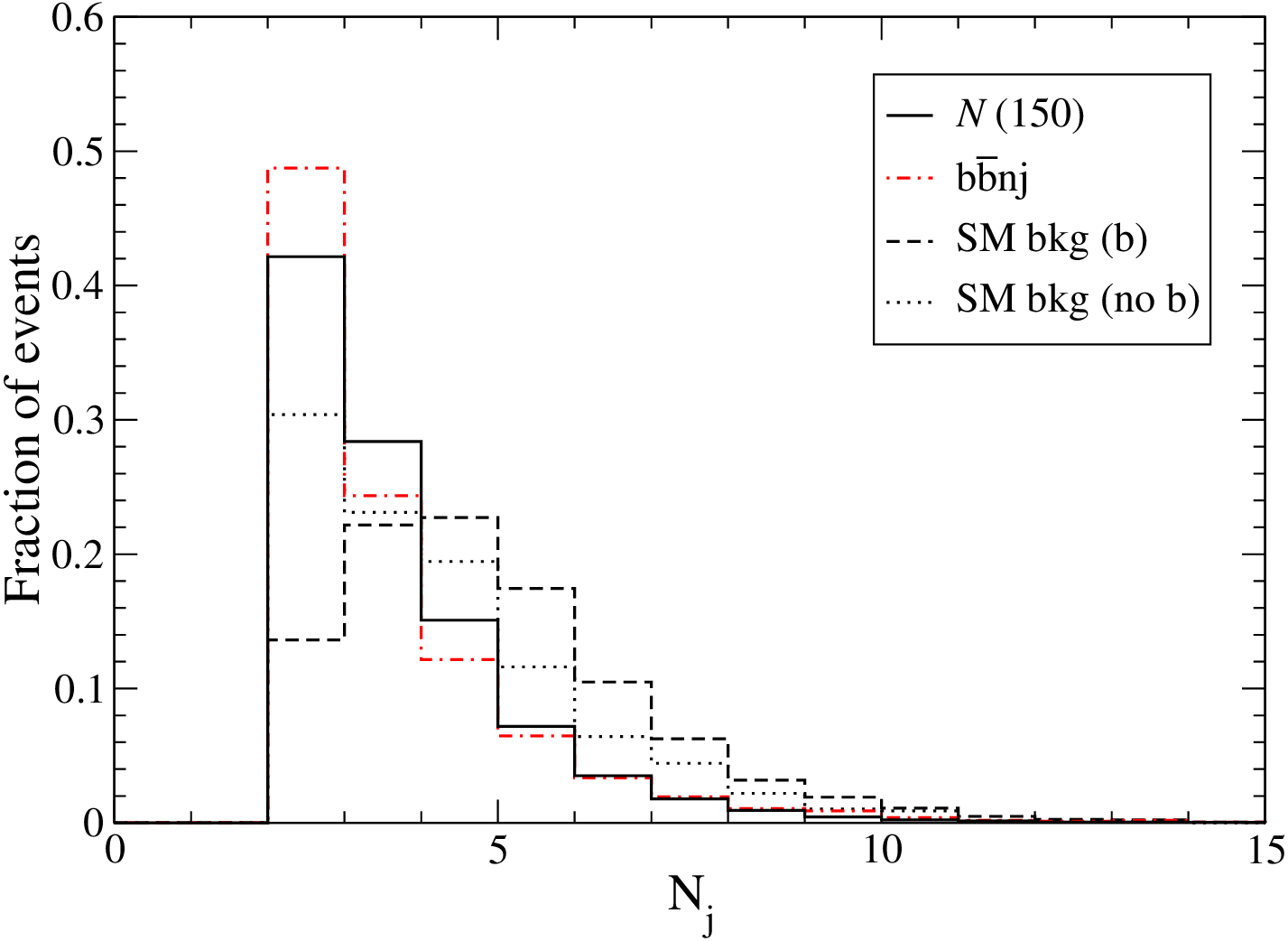,height=4.9cm,clip=} \\
\epsfig{file=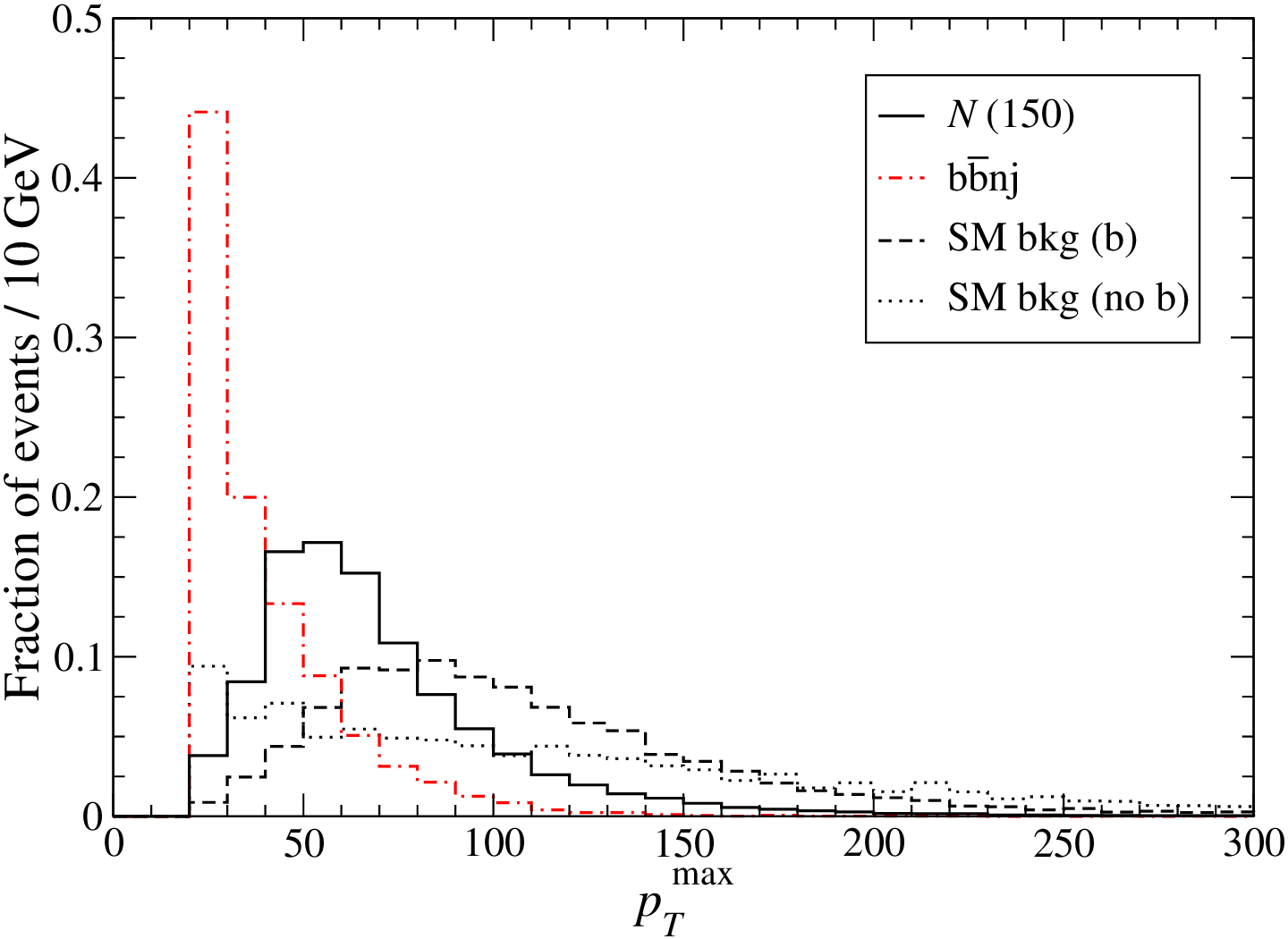,height=4.9cm,clip=} & &
\epsfig{file=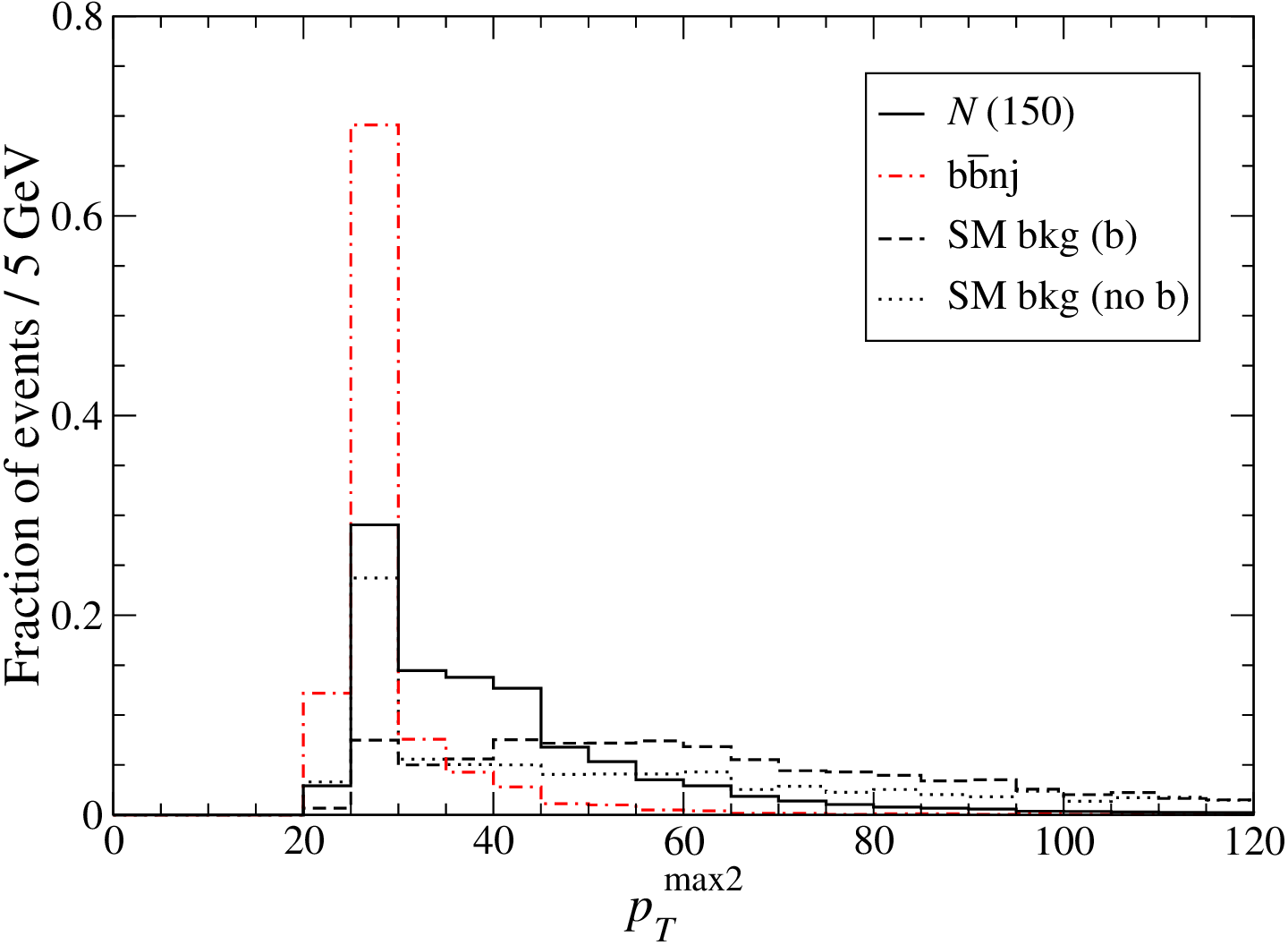,height=4.9cm,clip=}
\end{tabular}
\caption{Normalised distributions of several discriminating variables for the
$\mumu jj$ signal with $m_N = 150$ GeV and its backgrounds (see the text).}
\label{fig:vars2}
\end{center}
\end{figure}
These variables are plotted in Fig.~\ref{fig:var1} for the $\mumu jj$ signal
and the
backgrounds grouped in three classes with common features: (a) $b \bar b nj$,
where both muons come from $b$ quark decays (the contribution of $c \bar c nj$
is negligible); (b) $t \bar t nj$, $tj$ and
$W/Z b \bar b nj$, where one muon comes from a $b$ quark; (c) backgrounds
where both muons come from $W/Z$ decays (mainly di-boson and tri-boson
production). Kinematical cuts on the variables listed above do not render the
$\mu^\pm \mu^\pm jj$ final
state ``background free'', as it is apparent from the plots (and we have
explicitly checked). Indeed, for the large background cross sections in
Table~\ref{tab:Nsb150} the overlapping regions contain a large number of
background events, and they can be eliminated only by severely reducing the
signal. However, a likelihood analysis using
these and further variables can efficiently reduce the background. The
additional variables are shown in Fig.~\ref{fig:vars2}:
\begin{itemize}
\item The invariant mass $m_{jj}$ of the two jets with largest transverse
momentum, which for the signal are assumed to originate from the $W$ hadronic
decay, and the invariant mass of $\mu_2$ (the muon with lowest $p_T$) and these
two jets, $m_{W \mu_2}$. (Further details about the $W$ and $N$ mass
reconstruction can be found in appendix~\ref{sec:b}.) An important observation
in this case is that in backgrounds involving $b$ quarks this muon typically
has a small $p_T$, displacing  the background peaks to lower invariant masses.
\item The invariant mass of the two muons.
\item The separation between the muon with largest $p_T$ and the closest
jet, $\Delta R_{\mu_1 j}$. 
\item The number of $b$-tagged jets $N_b$ and jet multiplicities $N_j$.
Especially the former helps to separate the backgrounds involving 
$b$ quarks because they often have $b$-tagged jets. In this fast simulation
analysis we have fixed the
$b$-tagging efficiency to 60\%, but in a full simulation the $b$ tag probability
can be included in the likelihood function, improving the discriminating power
of this variable.
\item The transverse momenta of the two jets with largest
$p_T$, $p_T^\mathrm{max}$ and $p_T^\mathrm{max2}$ respectively.
\end{itemize}
These variables are not suited for performing kinematical cuts but greatly
improve the discriminating power of a likelihood function.
The resulting log-likelihood function is also shown in Fig. \ref{fig:logLS150}, 
where we distinguish four likelihood classes as in the previous figures: 
the signal, $b \bar b nj$, backgrounds with one muon from $b$ decays, 
and backgrounds with both muons from $W/Z$ decays.

\begin{figure}[htb]
\begin{center}
\epsfig{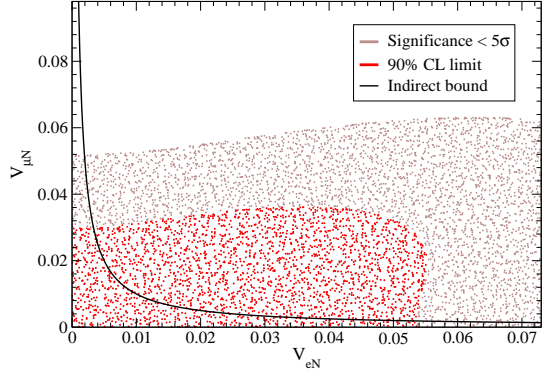} \\
\caption{Combined limits on $V_{eN}$ and $V_{\mu N}$, for $V_{\tau N} = 0$ and
$m_N = 150$ GeV. The red areas represent the 90\% CL limits if no signal is
observed. The white areas correspond to the region where a combined statistical
significance of $5\sigma$ or larger is achieved. The indirect limit from
$\mu-e$ LFV processes is also shown.}
\label{fig:comb150}
\end{center}
\end{figure}

The probability distributions built for $\mumu jj$ final states are used for
$\ee jj$ and $\mue jj$ as well.
As selection criteria we require $\log_{10} L_S/L_B \geq 1.4$ for $\mumu jj$
and $\log_{10} L_S/L_B \geq 2.5$ for $\ee jj$ and $\mue jj$ final states,
respectively, and that at least 
one of the two heavy neutrino mass assignments 
$m_{W \mu_1}$, $m_{W \mu_2}$ is between 130 and 170 GeV.\footnote{The latter
requirement assumes a previous knowledge of $m_N$. In the same way, the signal
distributions for the likelihood analysis must be built for a fixed $m_N$
value.
Thus, experimental searches must be performed by comparing data with
Monte-Carlo samples generated for different values of $m_N$. This procedure,
although more involved than a search with generic cuts, provides much better
sensitivity.}
The number of events surviving these cuts can be read on the right part of 
Table~\ref{tab:Nsb150}. As it is apparent, the likelihood analysis is
quite effective in suppressing backgrounds, especially $b \bar b nj$,
$t\bar t nj$ and $W/Z b \bar b nj$. 
The resulting
statistical significance for the heavy neutrino
signals are collected in Table~\ref{tab:sign150},
assuming a ``reference'' 20\% systematic uncertainty in the
backgrounds (which still has to be precisely evaluated in a dedicated study).
\begin{table}[htb]
\begin{center}
\begin{tabular}{cccc}
	        & $\mumu$      & $\ee$       & $\mue$ \\
$N~(\mathrm{a})$       & $16.2\sigma$ & $-$         & $-$  \\
$N~(\mathrm{b})$         & $-$          & $4.2\sigma$ & $-$  \\
$N~(\mathrm{c})$     & $11.4\sigma$ & $1.1\sigma$ & $5.5\sigma$ \\
\end{tabular}
\caption{Statistical significance of the heavy neutrino signals in the
different channels, for a mass $m_N = 150$ GeV and coupling (a) to the
muon, $V_{\mu N} = 0.098$; (b) to the electron, $V_{e N} = 0.073$; (c)
to both, $V_{e N} = 0.073$ and $V_{\mu N} = 0.098$.}
\label{tab:sign150}
\end{center}
\end{table}
The limits on heavy neutrino masses and couplings depend on the light lepton
they are coupled to. We can consider two extreme cases:
\begin{itemize}
\item[(a)] A 150 GeV heavy neutrino coupling only to the muon can be
discovered for mixings $|V_{\mu N}| \geq 0.054$, and if no background excess
is found the limits $|V_{\mu N}|^2 \leq 0.97\;(1.2) \times 10^{-3}$ can be set
at 90\% (95\%) CL, improving the ones from low energy processes (see
section~\ref{sec:2}) by a factor of 10. Heavy neutrino masses up to 200 GeV can
be observed with $5 \sigma$ at the LHC for $V_{\mu N} = 0.098$.
\item[(b)] A 150 GeV heavy neutrino coupling only to the electron can be
discovered for mixings $|V_{eN}| \geq 0.080$ (excluded by the limits in
section~\ref{sec:2}), but if no background excess is found the limits
$|V_{eN}|^2 \leq 2.1\;(2.5) \times 10^{-3}$, which are slightly better than
the one derived from
Eq.~(\ref{eps1}), can be set at 90\% (95\%) CL. Heavy neutrino masses up
to 145 GeV can be observed with $5 \sigma$ at the LHC for $V_{eN} = 0.073$.
\end{itemize}
For a heavy neutrino coupling to the electron and muon the limits depend on
both couplings as well as on its mass. The combined limits for $m_N = 150$ GeV
are displayed in Fig.~\ref{fig:comb150}. Except in the regions with $V_{eN}
\sim 0$ or $V_{\mu N} \sim 0$, the indirect limit from $\mu-e$ LFV processes,
also shown in this plot, is much more restrictive. 

These limits can be considered conservative in the sense that only the
lowest-order signal contribution (without hard extra jets at the partonic
level) has been included, and further signal contributions $\ell N nj$ should
improve the heavy neutrino observability. If the Higgs is heavier than 120 GeV
the branching ratios $\mathrm{Br}(N \to W \ell)$ will increase as well. We also
stress again that in the $\ee jj$ and $\mue jj$ channels the evaluation of
$t \bar t nj$ and other backgrounds with isolated electrons from $b$ quarks must
be confirmed with a full simulation, with an eventual optimisation of the
isolation criteria. This is beyond the scope of the present work.

It is worth explaining here in more detail why our results are much more
pessimistic than previous ones. With this purpose, we apply to signal
and backgrounds the sequential kinematical cuts in Ref.~\cite{Han:2006ip}:
\begin{itemize}
\item Missing energy $\ptmiss < 25$ GeV.
\item Lego-plot separation $\Delta R_{\mu j} > 0.5$.
\item Dijet invariant mass $60~\text{GeV} < m_{jj} < 100~\text{GeV}$, where
the two jets are expected to come from the $W$ boson in the case of the signal.
\end{itemize}
The number of events for the signal and main backgrounds after these cuts are
gathered in the left column of Table~\ref{tab:Nhan} (we do not show smaller
backgrounds for
brevity). For $m_N = 150$ GeV and $V_{\mu N} = 0.098$ the signal cross section
is reduced to 1.7 fb, to be compared to $\sim 2.2$ fb in
Ref.~\cite{Han:2006ip}.
But our total background cross section after cuts amounts to 44 fb, while
their estimate is of 0.04 fb.
This difference by a factor of $1000$ arises mainly from
the $b \bar b nj$ background, overlooked before,
which is by far the largest one. But even if $b \bar b nj$ is not taken into
account, the background cross section $\sim 0.88$ fb is 20 times larger,
due to:
(i) $t \bar t nj$, which was assumed negligible after cuts,
and $W/Z b \bar b nj$, also overlooked;
(ii) the $WZnj$ background, because parton-level analyses underestimate the
probability of missing a lepton and thus its contribution; 
(iii) pile-up, which makes lower order processes ($n < 2$) contribute.
All these backgrounds, collected in Table~\ref{tab:Nhan}, can be
compared to $WWW$, which was found to be the main
background before.
The resulting statistical significance of the
signal, ignoring systematic errors, is $S/\sqrt B = 1.41 \sigma$
for 30 fb$^{-1}$,
far from the $\sim 30 \sigma$ previously estimated.
(If one makes the more realistic assumption that systematic errors are of order
20\%, as we do in this work, then the statistical
significance is further reduced to $0.19\sigma$.) 
It could be argued that the cuts in the previous list might be strengthened in
order to further reduce the backgrounds. But this would be at the
cost of reducing the signal as well.
On the other hand, additional cuts on lepton transverse momenta can be
introduced to reduce $b \bar b nj$ and $t \bar t nj$. Requiring that one charged
lepton has $p_T \geq 30$ GeV and the other one $p_T \geq 20$ GeV, the signal is
hardly affected while $b \bar b nj$ is essentially eliminated, as it is shown
in the second column of Table~\ref{tab:Nhan}. The statistical significance in
this case is $S/\sqrt B = 14.1 \sigma$ 
(ignoring systematic errors) or $12.1 \sigma$ (with 20\% systematics).
We emphasise that, as it can be observed by comparing
Tables~\ref{tab:Nsb150} and \ref{tab:Nhan}, a probabilistic analysis is much
more powerful in this case than a standard one based on cuts.
But at any rate recovering parton-level estimates for the
signal significance seems hardly possible.  

\begin{table}[htb]
\begin{center}
\begin{tabular}{ccc}
& Sequential cuts I & Sequential cuts II \\
$N~(\mu)$       & 51.3     & 44.0 \\
$b \bar b nj$   & 1293     & 2.7  \\
$t \bar t nj$   & 15.3     & 1.4 \\
$W b \bar b nj$ & 3.6      & 0.2 \\
$W t \bar t nj$ & 0.7      & 0.7 \\
$Z b \bar b nj$ & 0.9      & 0.0 \\
$WW nj$         & 0.5      & 0.5 \\
$WZ nj$         & 4.1      & 2.9 \\
$WWW nj$        & 1.1      & 0.9 \\
\end{tabular}
\caption{Number of $\mu^\pm \mu^\pm jj$ events at LHC for 30 fb$^{-1}$,
after the kinematical cuts in Ref.~\cite{Han:2006ip} (first column) and with
additional requirements (second column, see the text). The heavy
neutrino signal is evaluated assuming $m_N = 150$ GeV and $V_{\mu N} = 0.098$.}
\label{tab:Nhan}
\end{center}
\end{table}

Finally, we would like to note that we have not addressed the observability of
heavy neutrino signals in $\tau$ lepton decay channels because they are
expected to have much worse sensitivity. For hadronic $\tau$ decays the charge
of the decaying lepton seems rather difficult to determine, hence backgrounds
from top pair and $Z$ production will be huge (see also section~\ref{sec:D}
below). For leptonic decays $\tau \to \ell \nu_\tau \bar \nu_{\ell}$,
$\ell=e,\mu$, not only the branching ratios are smaller,
but also the signal has final state neutrinos and thus the discriminating
power of $\ptmiss$ against di-boson and tri-boson backgrounds is much worse.

\subsection{$\ell^\pm \ell^\pm jj$ production for $m_N < M_W$}
\label{sec:60}

\begin{figure}[htb]
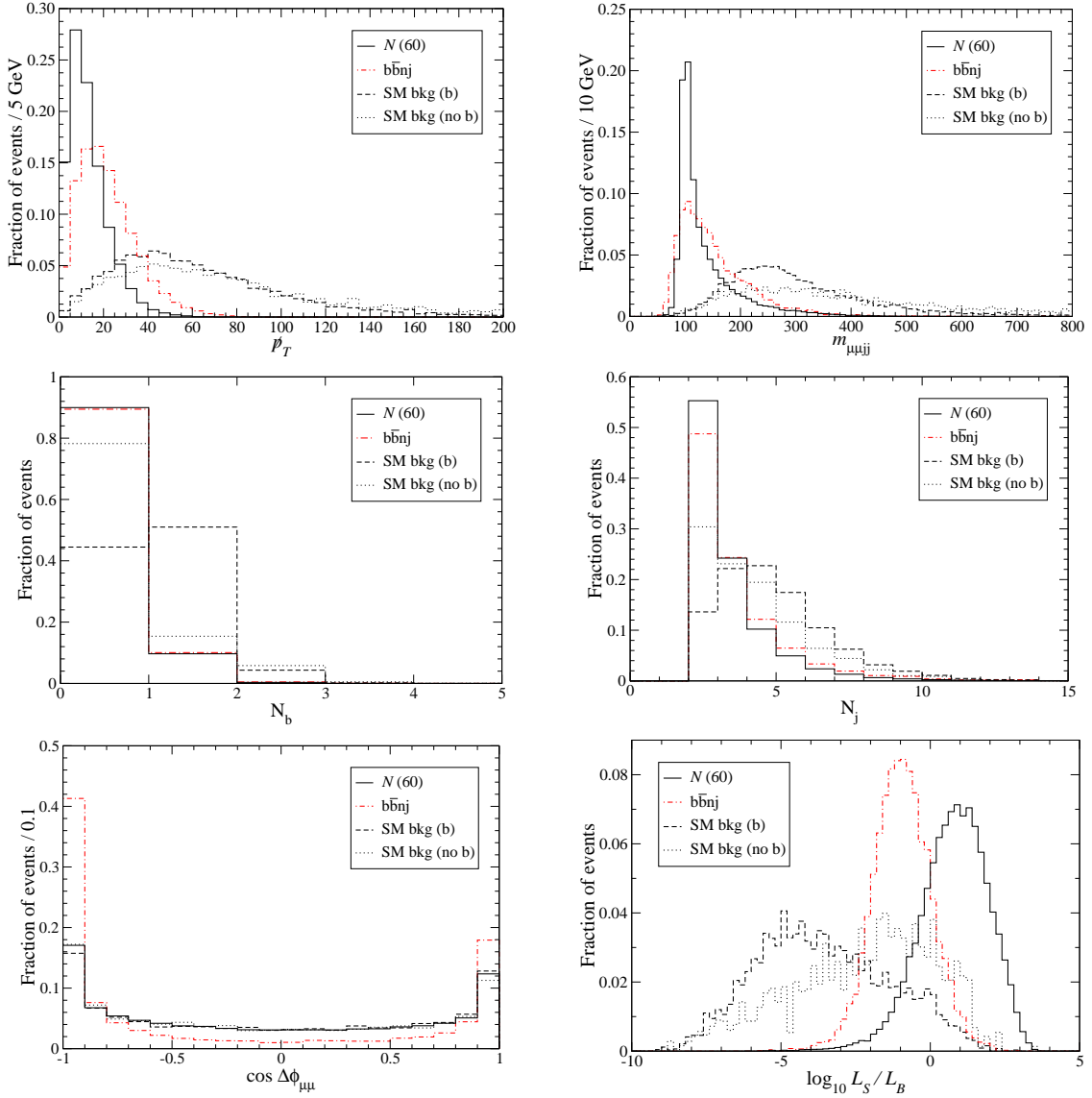

\begin{center}
\begin{tabular}{ccc}
\epsfig{file=./Figs/ptmiss-60.eps,height=4.9cm,clip=} & &
\epsfig{file=./Figs/Mlljj-60.eps,height=4.9cm,clip=} \\
\epsfig{file=./Figs/bmult-60.eps,height=4.9cm,clip=} & & 
\epsfig{file=./Figs/mult-60.eps,height=4.9cm,clip=} \\
\epsfig{file=Figs/cPhll-60.eps,height=4.9cm,clip=} & &
\epsfig{file=Figs/logLS-60.eps,height=4.9cm,clip=} \\
\end{tabular}
\caption{Normalised distributions of several discriminating variables for the
$m_N = 60$ GeV signal and the three background classes (see the
text). The last plot corresponds to the log-likelihood function.}
\label{fig:vars4}
\end{center}
\end{figure}

\begin{figure}[htb]
\begin{center}
\epsfig{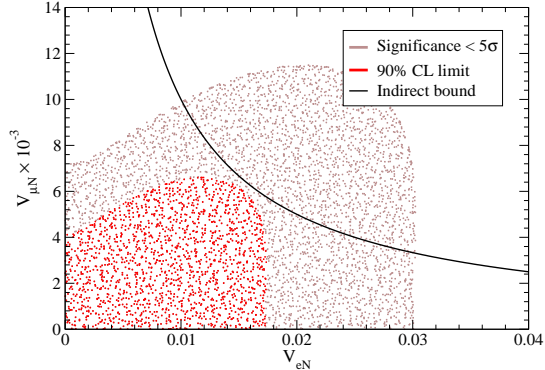} \\
\caption{Combined limits on $V_{eN}$ and $V_{\mu N}$, for $V_{\tau N} = 0$ and
$m_N = 60$ GeV. The red areas represent the 90\% CL limits if no signal is
observed. The white areas correspond to the region where a combined statistical
significance of $5\sigma$ or larger is achieved. The indirect limit from
$\mu-e$ LFV processes is also shown.}
\label{fig:comb60}
\end{center}
\end{figure}

In this mass region we take the reference values $m_N = 60$ GeV and
(a) $V_{\mu N} = 0.01$, $V_{eN} = V_{\tau N} = 0$; (b) $V_{e N} = 0.01$,
$V_{\mu N} = V_{\tau N} = 0$; (c) $V_{e N} = 0.01$, $V_{\mu N} = 0.01$,
$V_{\tau N} = 0$.
The pre-selection criteria are the same as before. The
likelihood analysis is performed distinguishing four classes:
the signal, $b \bar b nj$, backgrounds with one muon from $b$ decays, 
and backgrounds with both muons from $W/Z$ decays.
The relevant variables are depicted in Figs.~\ref{fig:vars3} and
\ref{fig:vars4}:
\begin{figure}[p]
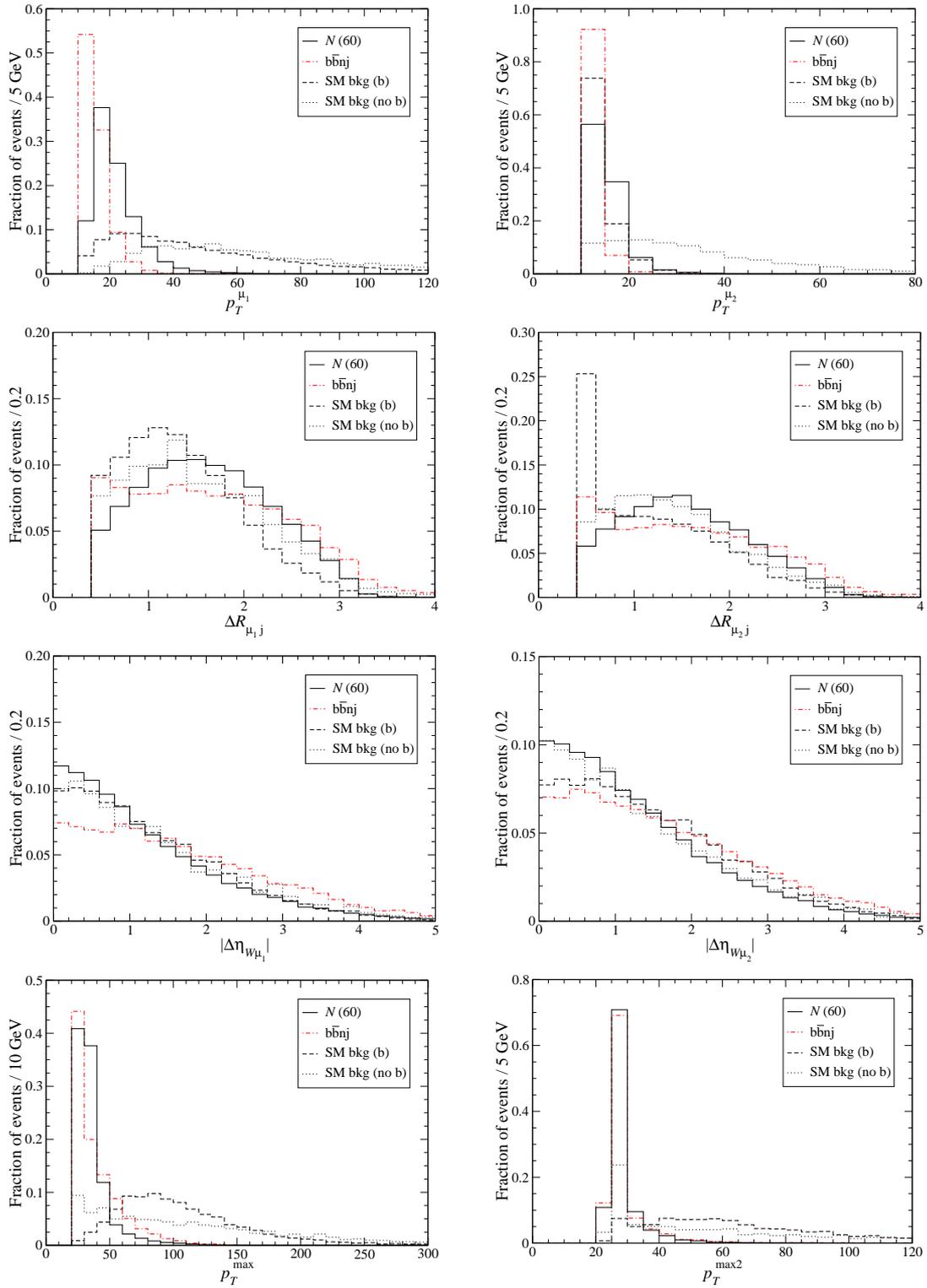

\begin{center}
\begin{tabular}{ccc}
\epsfig{file=Figs/ptl1-60.eps,height=4.9cm,clip=} & &
\epsfig{file=Figs/ptl2-60.eps,height=4.9cm,clip=} \\
\epsfig{file=Figs/dRl1j-60.eps,height=4.9cm,clip=} & &
\epsfig{file=Figs/dRl2j-60.eps,height=4.9cm,clip=} \\
\epsfig{file=./Figs/dhWl1-60.eps,height=4.9cm,clip=} & & 
\epsfig{file=./Figs/dhWl2-60.eps,height=4.9cm,clip=} \\
\epsfig{file=./Figs/ptmax-60.eps,height=4.9cm,clip=} & &
\epsfig{file=./Figs/ptmax2-60.eps,height=4.9cm,clip=} \\
\end{tabular}
\caption{Normalised distributions of several discriminating variables for the
$m_N = 60$ GeV signal and the three background classes (see the text).}
\label{fig:vars3}
\end{center}
\end{figure}
\begin{itemize}
\item The transverse momenta of the two muons (slightly smaller for $b \bar b
nj$ than for the signal, and much larger for the other backgrounds).
\item The distance between them and the closest jet, which is a good
discriminator against $t \bar t nj$ but not against $b \bar b nj$.
\item The rapidity difference between the muons and the $W^*$
from $N$ decay, which is reconstructed from the two jets with highest $p_T$.
\item The transverse momenta of the two jets with largest $p_T$. Again, these
two variables are excellent discriminators against high-$p_T$ backgrounds like
$t \bar t nj$ and di-boson production, but not very useful for $b \bar b nj$. 
\item The missing transverse momentum.
\item The invariant mass of the two muons and two jets with highest $p_T$,
$m_{\mu \mu jj}$. For
the signal, these four particles result from the decay of an on-shell $W$ boson,
so the distribution is very peaked around 100 GeV (the position of the peak is
displaced as a consequence of pile-up, which generates jets with larger
$p_T$ than the ones from the signal itself). Unfortunately, for
$b \bar b nj$ the distribution is quite similar.
\item The number of $b$ tags and the jet multiplicity.
\item The azimuthal angle (in transverse plane) between the two muons,
$\phi_{\mu \mu}$. For $b \bar b$
without additional jets this angle is often close to $180^\circ$, but for
$b \bar b j$ and higher order processes (which are also huge) this no longer
holds.
\end{itemize}
The resulting log-likelihood function is presented in Fig.~\ref{fig:vars4}. As
it can be easily noticed with a quick look at the variables presented, the
kinematics of $b \bar b nj$ is
very similar to the signal and so this background is very difficult to
eliminate. In particular, for larger $m_N$ requiring large transverse momentum
for the leptons drastically reduces $b \bar b nj$
(as seen in the previous subsection),
but for $m_N < M_W$ it reduces significantly the signal as well. 
As selection cut we require $\log_{10} L_S / L_B \geq 2.2$ for the three final
states, which practically eliminates all backgrounds except $b \bar b nj$.
The number of remaining background events is given in the right
part of Table~\ref{tab:Nsb60} (numbers of background events at pre-selection
equal those in Table~\ref{tab:Nsb150}, and are quoted on the left for
better comparison).
\begin{table}[htb]
\begin{center}
\begin{tabular}{cccccccc}
& \multicolumn{3}{c}{Pre-selection} & \hspace{.5cm} &
  \multicolumn{3}{c}{Selection} \\
                 & $\mumu$  & $\ee$    & $\mue$ 
	     &   & $\mumu$  & $\ee$    & $\mue$ \\
$N~(\mathrm{a})$ & 427.3    & 0        & 0 
             &   & 42.1     & 0        & 0  \\
$N~(\mathrm{b})$         & 0        & 174.7    & 0 
             &   & 0        & 33.9     & 0  \\
$N~(\mathrm{c})$ & 214.0    & 88.5     & 290.9 
             &   & 20.4     & 17.1     & 39.3 \\
$b \bar b nj$    & 14800    & 52000    & 82000 
             &   & 10.7     & 291      & 96  \\
$c \bar c nj$    & (11)     & 300      & 200 
             &   & (0)      & 0        & 0   \\
$t \bar t nj$    & 1162.1   & 8133.0   & 15625.3
             &   & 0.3      & 1.3      & 1.3 \\
$tj$             & 60.8     & 176.5    & 461.5
             &   & 0.0      & 0.0      & 0.1 \\
$W b \bar b nj$  & 124.9    & 346.7    & 927.3
             &   & 0.2      & 2.4      & 1.3 \\
$W t \bar t nj$  & 75.7     & 87.2     & 166.9
             &   & 0.0      & 0.0      & 0.0 \\
$Z b \bar b nj$  & 12.2     & 68.9     & 117.0
             &   & 0.0      & 1.4      & 0.2 \\
$WW nj$          & 82.8     & 89.0     & 174.8
             &   & 0.0      & 0.0      & 0.0 \\
$WZ nj$          & 162.4    & 252.0    & 409.2
             &   & 0.6      & 0.4      & 0.5 \\
$ZZ nj$          & 3.8      & 13.3     & 12.9
             &   & 0.0      & 0.5      & 0.1 \\
$WWW nj$         & 31.9     & 30.1     & 64.8
             &   & 0.9      & 0.0      & 0.0    
\end{tabular}
\caption{Number of $\ell^\pm \ell^\pm jj$ events at LHC for 30 fb$^{-1}$, at the
pre-selection and selection levels. The heavy
neutrino signal is evaluated assuming $m_N = 60$ GeV and coupling
(a) to the
muon, $V_{\mu N} = 0.01$; (b) to the electron, $V_{e N} = 0.01$; (c)
to both, $V_{e N} = 0.01$ and $V_{\mu N} = 0.01$.}
\label{tab:Nsb60}
\end{center}
\end{table}
Requiring larger $L_S/L_B$ for the $\ee jj$ and $\mue jj$ channels does not
improve the results, because it decreases the signals too much. The resulting
statistical significance for the heavy neutrino signals are collected in
Table~\ref{tab:sign60}, assuming a 20\% systematic uncertainty in the
backgrounds.
\begin{table}[htb]
\begin{center}
\begin{tabular}{cccc}
	        & $\mumu$      & $\ee$        & $\mue$ \\
$N~(\mathrm{a})$       & $10.0\sigma$ & $-$          & $-$  \\
$N~(\mathrm{b})$         & $-$          & $0.54\sigma$ & $-$  \\
$N~(\mathrm{c})$     & $4.9\sigma$  & $0.28\sigma$ & $1.75\sigma$ \\
\end{tabular}
\caption{Statistical significance of the heavy neutrino signals in the
different channels, for a mass $m_N = 60$ GeV and coupling
(a) to the
muon, $V_{\mu N} = 0.01$; (b) to the electron, $V_{e N} = 0.01$; (c)
to both, $V_{e N} = 0.01$ and $V_{\mu N} = 0.01$.}
\label{tab:sign60}
\end{center}
\end{table}
From these significances, the following limits can be extracted:
\begin{itemize}
\item[(a)] A 60 GeV neutrino coupling only to the muon can be
discovered for mixings $|V_{\mu N}| \geq 0.0070$, and bounds
$|V_{\mu N}|^2 \leq
1.65 (1.95) \times 10^{-5}$ can be set at 90\% (95\%) CL if a background
excess is
not observed. These figures are $\sim 35$ times worse than in previous
parton-level estimates which overlooked the main background
$b \bar b nj$, but would still improve the direct limit from L3
\cite{Adriani:1992pq,Achard:2001qv} by an order of magnitude.
\item[(b)] A 60 GeV neutrino coupling only to the electron can be
discovered for mixings $|V_{e N}| \geq 0.030$, and bounds
$|V_{eN}|^2 \leq
3.1 (3.6) \times 10^{-4}$ can be set at 90\% (95\%) CL if a background excess
is not observed.
\end{itemize}
The general limits for a heavy neutrino coupling to the electron and muon
are displayed in Fig.~\ref{fig:comb60}. It is interesting to observe that the
direct limit from non-observation of like-sign dileptons at LHC will be more
restrictive than indirect ones from $\mu-e$ LFV processes at low energies.

\subsection{Opposite-sign dilepton signals}
\label{sec:D}

In final states $e^\pm \mu^\mp jj$ the analysis is similar but the
backgrounds are much larger. In particular, opposite-sign lepton pairs from
$b \bar b nj$ production are much more abundant than like-sign pairs.
Opposite-sign dileptons are produced from $t \bar t nj$ dileptonic decays and
$W^+ W^- nj$ production (which is larger than $W^\pm W^\pm nj$).
We assume a heavy Dirac neutrino with a mass of 60 GeV and
$V_{eN} = 0.01$, $V_{\mu N} = 0.01$. A Majorana neutrino gives this signal too,
but with half the cross section for the same couplings. We
use the same pre-selection cuts as in the like-sign dilepton analysis but
requiring instead opposite charge for the leptons. 
The number of signal and background events at pre-selection is collected in the
left column of Table~\ref{tab:Nsb60D}. At pre-selection the $b \bar b nj$,
$t \bar t nj$ and $WWnj$ backgrounds are 7, 15 and 70 times larger,
respectively, than
the corresponding ones for $\mu^\pm e^\mp$ in Table~\ref{tab:Nsb60}.

\begin{table}[htb]
\begin{center}
\begin{tabular}{cccc}
& Pre-selection & \hspace{.5cm} & Selection \\
$N~(e,\mu)$     & 593.5    & & 14.7\\
$b \bar b nj$   & 602000   & & 73 \\
$c \bar c nj$   & 5750     & & 0 \\
$t \bar t nj$   & 233135.1 & & 0.3 \\
$tj$            & 1003.8   & & 0.0  \\
$W b \bar b nj$ & 927.6    & & 0.0 \\
$W t \bar t nj$ & 197.0    & & 0.0 \\
$Z b \bar b nj$ & 180.8    & & 0.0 \\
$WW nj$         & 12016.5  & & 0.7  \\
$WZ nj$         & 412.1    & & 0.0 \\
$ZZ nj$         & 14.2     & & 0.0 \\
$WWW nj$        & 131.4    & & 0.0
\end{tabular}
\caption{Number of $\mu^\pm e^\mp jj$ events at LHC for 30 fb$^{-1}$, at the
pre-selection and selection levels. The heavy
neutrino signal is evaluated assuming $m_N = 60$ GeV and coupling to electron
and muon $V_{e N} = 0.01$, $V_{\mu N} = 0.01$.}
\label{tab:Nsb60D}
\end{center}
\end{table}

\begin{figure}[p]
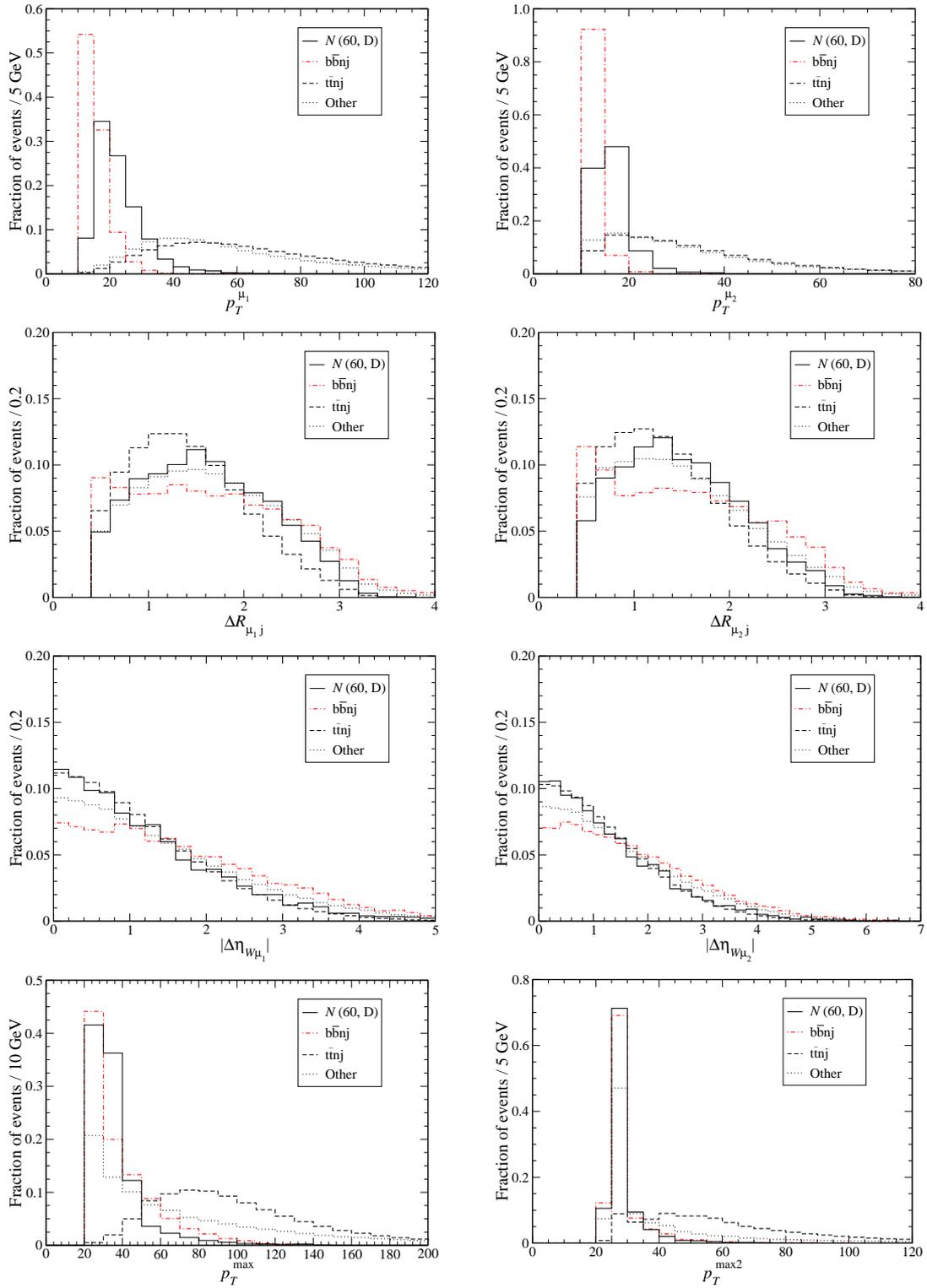

\begin{center}
\begin{tabular}{ccc}
\epsfig{file=Figs/ptl1-60D.eps,height=4.9cm,clip=} & &
\epsfig{file=Figs/ptl2-60D.eps,height=4.9cm,clip=} \\
\epsfig{file=Figs/dRl1j-60D.eps,height=4.9cm,clip=} & &
\epsfig{file=Figs/dRl2j-60D.eps,height=4.9cm,clip=} \\
\epsfig{file=./Figs/dhWl1-60D.eps,height=4.9cm,clip=} & & 
\epsfig{file=./Figs/dhWl2-60D.eps,height=4.9cm,clip=} \\
\epsfig{file=./Figs/ptmax-60D.eps,height=4.9cm,clip=} & &
\epsfig{file=./Figs/ptmax2-60D.eps,height=4.9cm,clip=} \\
\end{tabular}
\caption{Normalised distributions of several discriminating variables for a
60 GeV Dirac neutrino and the three background classes (see the text).}
\label{fig:vars5}
\end{center}
\end{figure}
\begin{figure}[p]
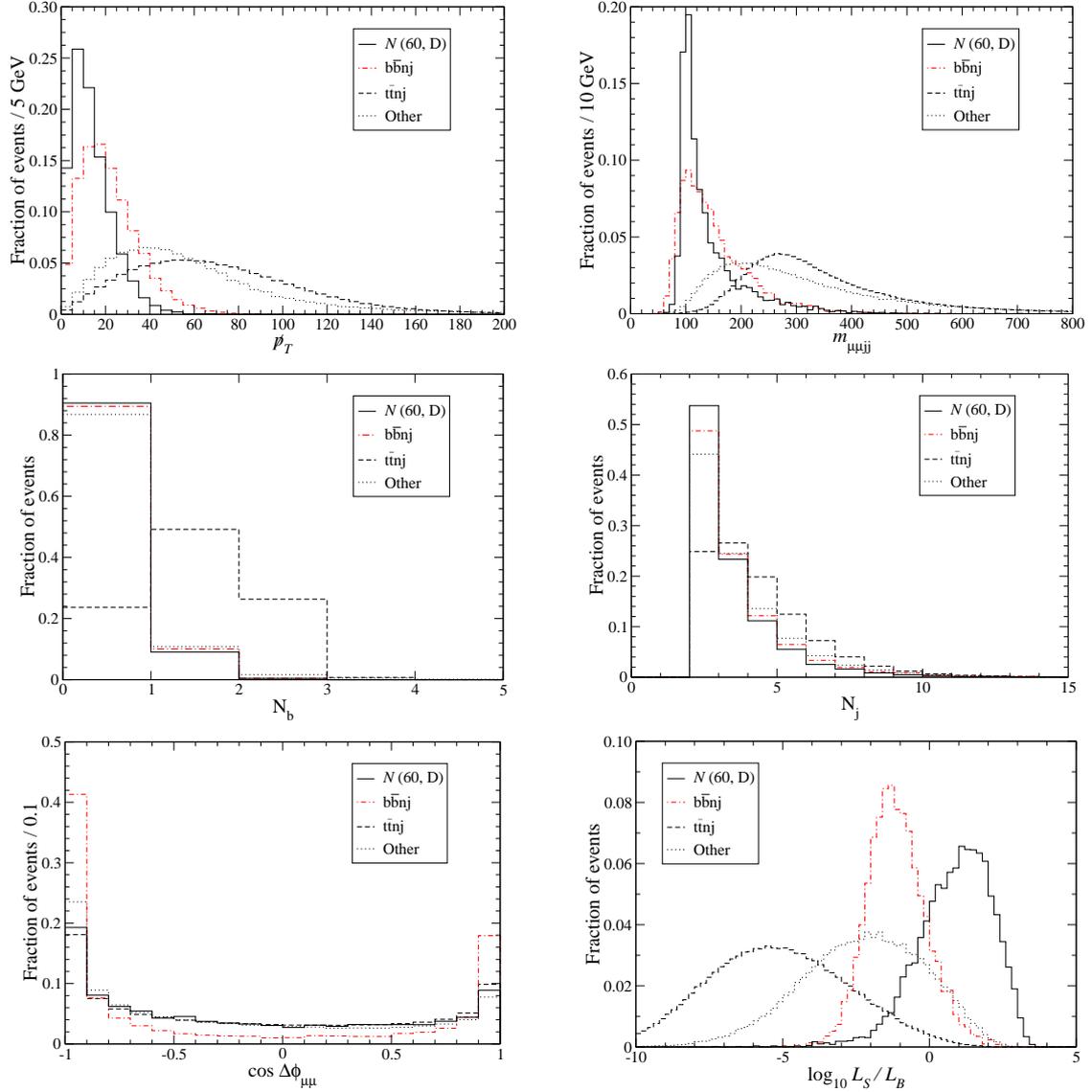

\begin{center}
\begin{tabular}{ccc}
\epsfig{file=./Figs/ptmiss-60D.eps,height=4.9cm,clip=} & &
\epsfig{file=./Figs/Mlljj-60D.eps,height=4.9cm,clip=} \\
\epsfig{file=./Figs/bmult-60D.eps,height=4.9cm,clip=} & & 
\epsfig{file=./Figs/mult-60D.eps,height=4.9cm,clip=} \\
\epsfig{file=Figs/cPhll-60D.eps,height=4.9cm,clip=} & &
\epsfig{file=Figs/logLS-60D.eps,height=4.9cm,clip=} \\
\end{tabular}
\caption{Normalised distributions of several discriminating variables for a
60 GeV Dirac neutrino and the three background classes (see the
text). The last plot corresponds to the log-likelihood function.}
\label{fig:vars6}
\end{center}
\end{figure}

\begin{figure}[htb]
\begin{center}
\epsfig{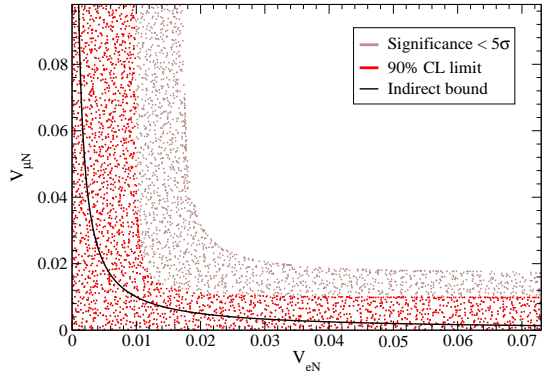} \\
\caption{Combined limits on $V_{eN}$ and $V_{\mu N}$ for a 60 GeV Dirac
neutrino. The red areas represent the 90\% CL limits if no signal is observed. 
The white areas correspond to the region where a combined statistical
significance of $5\sigma$ or larger is achieved. The indirect limit from
$\mu-e$ LFV processes is also shown.}
\label{fig:comb60D}
\end{center}
\end{figure}

The kinematical variables useful for discriminating the signal against
the backgrounds are the same as for a 60 GeV heavy Majorana neutrino in the
like-sign
dilepton channels. However, in this case the distributions for some
backgrounds, namely $t \bar t nj$ and $WWnj$, are different. We have grouped
backgrounds in three classes: $b \bar b nj$,  $t \bar t nj$, and the other
backgrounds
(dominated by $WWnj$). The distributions for the relevant kinematical
variables and the log-likelihood function are collected in
Figs.~\ref{fig:vars5} and \ref{fig:vars6}. For event selection we require
$\log_{10} L_S / L_B \geq 2.9$, yielding the number of events in the right
columns of Table~\ref{tab:Nsb60D}.
The significance of the heavy Dirac neutrino signal is only
$0.86\sigma$. The combined limits on $V_{eN}$ and $V_{\mu N}$ are presented
in Fig.~\ref{fig:comb60D}. The shape of the regions is drastically different
from Figs.~\ref{fig:comb150} and \ref{fig:comb60} because the sensitivity in
the $e^+ e^- jj$ and $\mu^+ \mu^- jj$ channels is negligible, and only when
$N$ couples sizeably to both electron and muon the heavy neutrino signal is
statistically significant in the $\mu^\pm e^\mp jj$ channel.
The direct limit from non-observation of a $\mu^\pm e^\mp jj$ excess has a
similar shape as the indirect limit but it is less restrictive in all cases.

\section{Estimates for Tevatron}
\label{sec:5}

The observability of heavy neutrino signals in like-sign
dilepton channels at Tevatron seems to be dominated by the size of the signal
itself. In contrast with LHC, backgrounds are much smaller. For example, the
$WZjj$ and $WWjj$ backgrounds have cross sections of 0.1 and 0.09 fb,
respectively, with the cuts in Eq.~(\ref{ec:gcuts}). Then,
it seems reasonable to estimate the
total background for 1 fb$^{-1}$ (including $b \bar b$) as one event. This
rough estimation is in agreement with the detailed calculation in
Ref.~\cite{D0},
in which $b \bar b$ is estimated from real data.
Therefore, if signal events have
not been observed with the already collected luminosity, upper limits of
3.36 and 4.14 events  \cite{Feldman:1997qc} can be set on the signal, at 90\%
and 95\% CL, respectively. From Fig.~\ref{fig:cross},
and for a fixed mass $m_N = 60$ GeV,
this implies upper bounds $|V_{\mu N}|^2 \leq 1.3 \times 10^{-4}$ (90\% CL)
$|V_{\mu N}|^2 \leq 1.6 \times 10^{-4}$ (95\% CL). This would slightly improve
the limits from L3~\cite{Adriani:1992pq,Achard:2001qv}.
Of course, a detailed simulation with the already collected data is
necessary to make any claim, and the limits eventually obtained will depend on
the actual number of observed like-sign dilepton events.

Note also that, given the cross sections in Fig.~\ref{fig:cross}, for a
luminosity of 1 fb$^{-1}$ and $V_{\mu N} = 0.098$ the heavy neutrino signals
only exceed a handful of events for heavy neutrino masses $m_N < M_W$, and
thus the Tevatron sensitivity (when acceptance and efficiency are taken into
account)
is limited to this mass range. This also holds for a heavy neutrino mixing with
the tau lepton, for which the $N$ production can be larger but $\tau$ decay
branching fractions must be also included in the final cross section.
Then, if the small excess found by CDF
\cite{Abulencia:2007rd} is confirmed, its explanation through heavy neutrinos
requires additional interactions, for example mediated by a new $Z'$ boson
\cite{delAguila:2007ua}.

\section{Conclusions}
\label{sec:6}

Large hadron colliders are not in principle the best place to search for
new heavy neutral leptons. However, Tevatron is performing quite well and LHC
will start operating soon, so one must wonder if the large
electroweak rates available at large hadron colliders allow to discover new
heavy neutrinos, given the present constraints on them,
or improve these constraints. This is indeed the case in models with extra
interactions \cite{Ferrari:2000sp,Gninenko:2006br,Datta:1992qw}.
In this work we have, however, assumed that no other interactions exist and
that heavy neutrinos couple to the SM particles through its small mixing with
the known leptons.

Heavy Dirac or Majorana neutrinos with a significant coupling to the electron
can be best produced and seen at $e^+ e^-$ colliders in $e^+ e^- \to N \nu \to
\ell W \nu$,
which has a large cross section and whose backgrounds have a moderate size
\cite{Azuelos:1993qu,Gluza:1996bz,delAguila:2005mf,delAguila:2005pf}. On the
contrary, a Majorana $N$ mainly coupling to the muon is easier to discover at
a hadronic machine like LHC, in the process $q \bar q' \to W^+ \to \mu^+ N$
with subsequent decay $N \to \mu^+ W \to \mu^+ q \bar q'$
(plus the charge conjugate).
However, even this LNV final state is not easy to deal with. SM
backgrounds are large and require a careful analysis, especially
those involving $b$ quarks, for example $b \bar b nj$ and $t \bar t n j$ which
are the largest ones.

For the simulation of the $\ell^\pm \ell^\pm jj$ signals from heavy neutrinos
we have implemented heavy neutrino production in the ALPGEN framework. In the
$\mumu jj$ channel
we have shown, using a fast detector simulation,
that a hevy neutrino with a mixing $V_{\mu N} = 0.098$ can be
discovered with a $5\sigma$ significance up to masses $m_N = 200$ GeV.
In the region $m_N < M_W$ we find that a 60 GeV neutrino can be
discovered for mixings $|V_{\mu N}| \geq 0.0070$; upper limits
$|V_{\mu N}|^2 \leq
1.65 (1.95) \times 10^{-5}$ can be set at 90\% (95\%) CL if a $\mumu jj$
excess is
not observed. These figures are in sharp constrast with previous estimates,
and correspond to the increase in the background estimation of about two
orders of magnitude (three for $m_N < M_W$).
In particular, special care has to
be taken with $b \bar b$ plus jets. The probability of a $b \bar b$ pair to give
two like-sign isolated muons is tiny, but on the other hand the $b \bar b$
cross section $\sim 1~\mu$b is huge. A reliable background calculation requires
solving this $0 \cdot \infty$ indetermination, what is a
computationally very demanding task in which some apparently reasonable
simplifying  assumptions, like requiring high transverse momenta of $b$ quarks
at generator level, can result in an underestimation by a factor of 30.
The $b \bar b nj$ background has been found to be negligible for
larger $m_N$ values but dominant for $m_N < M_W$
(after cuts, 5 times larger than the sum of the other backgrounds).
This behaviour is due
to the very different signal kinematics in these two cases.
For $m_N < M_W$ the charged leptons are produced with very small transverse
momentum, therefore a cut on this variable, which could be efficiently used
to remove $b \bar b nj$, cannot be applied. On the other hand,
requiring {\em e.g.} that one charged lepton has $p_T > 30$ GeV and the other
one $p_T > 20$ GeV hardly affects the signal for $m_N = 150$ GeV, while it
practically eliminates $b \bar b nj$.

For the other like-sign dilepton channels, $\ee jj$ and $\mue jj$,
the prospects are worse because backgrounds are much larger.
We have found that
a heavy neutrino with $V_{eN} = 0.073$ can be discovered 
with $5\sigma$ up to masses $m_N = 145$ GeV. In the region
$m_N < M_W$, a heavy neutrino with $m_N = 60$ GeV can be discovered 
for mixings $|V_{eN}| \geq 0.030$; upper limits
$|V_{eN}|^2 \leq 3.1 (3.6) \times 10^{-4}$ can be set at 90\% (95\%) CL if a
background excess is
not observed. The latter limits are of the same magnitude but worse than
those from L3. Besides, couplings of this size
would be in conflict with the non-observation of neutrinoless double beta
decay, requiring cancellations with other new physics contributions. Finally,
for a heavy neutrino with $m_N = 60$ GeV and coupling to both electron and muon
we have found that
direct limits on $V_{eN}$ and $V_{\mu N}$ will improve the indirect ones
from $\mu-e$ LFV processes. For completeness we have also examined the LHC
sensitivity for a Dirac neutrino coupling to the electron and muon,
in $\mu^\pm e^\mp jj$ final states. The sensitivity is much worse, as expected
from the larger LNC backgrounds involving opposite-sign dileptons, and the
direct limits obtained are worse than the present indirect ones.
Hence, LHC is not expected to provide any useful
direct limit on heavy Dirac neutrinos, for which all final states conserve
lepton number. By the same token, other decay channels such as $N \to
Z\nu$, $N \to H \nu$ and production processes as $pp \to Z \to N \nu$,
have too large backgrounds as well.

In the detailed analyses presented for $m_N = 150$ GeV and $m_N = 60$ GeV
we have shown that background suppression ($t \bar t nj$ and diboson production
in the former case, $b \bar b nj$ in the latter) is not efficient
with simple kinematical cuts, and requires more sophisticated methods, like the
likelihood analysis applied here, or neural networks. The analysis could be
further improved when one includes other variables not accessible at the level
of fast simulation. For example, a $b \bar b$ pair giving two like-sign
isolated muons is most often caused by the oscillation of one of the $B^0$
mesons before decay.
This should appear as a secondary vertex, which could be identified in the
detector. On the other hand, the possibility of lepton charge misidentification
should be addressed.
Full simulation of $b \bar b nj$ for the LHC luminosity is beyond
reach of present and foreseable computers, so this background will have to be
estimated from data. In any case, we stress that  $b \bar b nj$, as well as $t
\bar t nj$, must always be considered as a potentially dangerous source of two
like-sign dileptons. And, if a moderate background excess is observed
at low transverse momenta, a precise evaluation of the $b \bar b nj$
background is compulsory before drawing any conclusion.

It is finally worth noting that heavy neutrino decays, as for any other
vector-like fermion, are a source of Higgs bosons \cite{delAguila:1989rq}.
Nevertheless, in contrast with the quark sector \cite{Aguilar-Saavedra:2006gw}
Higgs boson production from $N$ decays is rather small, and only a handful
of $\mu N \to \mu \nu H \to \mu \nu b \bar b$ events are expected to be found
at LHC. Besides, we also point out that large effects due to heavy
neutrinos and,
more generally, other neutrino physics beyond the SM might be observed at
large hadron colliders. However, in all cases they require new interactions
and often model dependent constraints. This means further assumptions, and
in this situation the main novel ingredient is not only the heavy neutrino.
In this category there are many interesting scenarios, also including
supersymmetry (see for an example Refs.~\cite{Porod:2000hv,Hirsch:2003fe}).
Then, compared to these new physics models the limits established in this work
are modest. For example, if the heavy neutrino has an interaction with a
typical gauge strength, as in left-right models with a new
heavy $W_R$, the LHC reach for $m_N$ increases up to approximately 2 TeV
\cite{Ferrari:2000sp,Gninenko:2006br}. In the case of a new leptophobic
$Z'_\lambda$
boson, the $m_N$ reach in $N$ pair production $pp \to Z'_\lambda \to NN$
is increased up to 800 GeV \cite{delAguila:2007ua}.

\vspace{1cm}
\noindent
{\Large \bf Acknowledgements}
\vspace{0.3cm}

\noindent
This work has been supported by MEC project FPA2006-05294,
Junta de Andaluc{\'\i}a projects FQM 101 and FQM 437, 
MIUR under contract 2006020509\_004,
and by the European Community's Marie-Curie Research Training
Network under contract MRTN-CT-2006-035505 ``Tools and Precision
Calculations for Physics Discoveries at Colliders''.
J.A.A.-S. acknowledges support by a MEC Ram\'on y Cajal contract.

\appendix
\section{Evaluation of the $b \bar b$ background}
\label{sec:a}

$b \bar b$ production, which has a huge cross section of order 1 $\mu$b at LHC,
is the largest SM source of like-sign dileptons. Charged leptons are produced
in the decays $b \to c \ell^- \nu$, $\bar b \to \bar c \ell^+ \nu$, and
like-sign lepton pairs can arise when one of the
$b$ quarks yields a $B^0$ meson which oscillates before decay. Additionally,
like-sign
charged leptons can be produced from the subsequent decay of the charm
(anti)quark, {\em e.g.} $c \to s \ell^+ \nu$. We have investigated the relative
contribution of the two
sources by simulating with {\tt Pythia} a $b \bar b$ sample of 25 million
events with and without $B^0$ mixing. The number of dielectron and dimuon
events (requiring isolation and transverse momentum greater than 10 GeV) is
gathered in Table~\ref{tab:Bmix}. A quick look at these numbers reveals that
about $80\%$ of like-sign dileptons result from $B^0$ oscillation.

\begin{table}[htb]
\begin{center}
\begin{tabular}{lcc}
& $B^0$ mixing & No $B^0$ mixing \\
\hline
$\mu^\pm \mu^\pm$ & 55   & 12 \\
$\mu^\pm e^\pm$   & 456  & 109 \\ 
$e^\pm e^\pm$     & 1242 & 334 \\
$\mu^+ \mu^-$     & 309  & 335 \\
$\mu^\pm e^\mp$   & 1357 & 1643 \\ 
$e^\pm e^\pm$     & 3755 & 4671 \\
\end{tabular}
\caption{Number of dilepton events obtained from a sample of 25 million
$b \bar b$ events, when $B^0$ oscillation is allowed in
{\tt Pythia} (first column) or not (second column).}
\label{tab:Bmix}
\end{center}
\end{table}

It must be emphasised that the theoretical evaluation of the $b \bar b$
contribution  to the like-sign dilepton SM background involves several
uncertainties. The most obvious one affects the total $b \bar b$ cross section,
which depends to a large extent on the generation cuts placed on $b$ transverse
momenta. A second one involves $b$ quark fragmentation. We have used the
Peterson parameterisation with $\epsilon_b = 0.0035$  \cite{Barate:1996fi}.
With the default {\tt Pythia} setting $\epsilon_b = 0.005$ the number of
(isolated) dileptons obtained is a factor $\sim 0.77$ smaller. But perhaps the
largest uncertainty comes from the fact that our analysis relies on a fast
simulation of the detector, which may be inadequate when studying delicate
issues like lepton isolation. At any rate, a full simulation of a large
$b \bar b$ sample is out of present reach and this background will have to be
measured using real data.

\begin{figure}[htb]
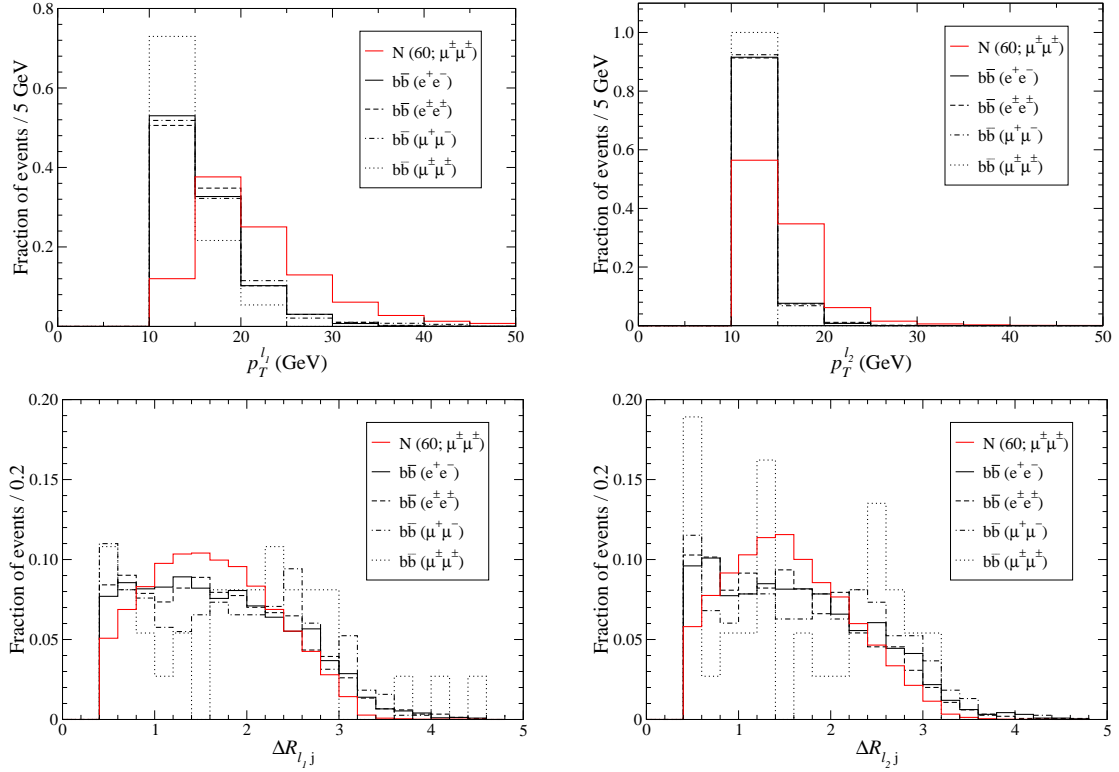

\begin{center}
\begin{tabular}{ccc}
\epsfig{file=Figs/PTl1-BB.eps,height=5cm,clip=} & \quad &
\epsfig{file=Figs/PTl2-BB.eps,height=5cm,clip=} \\
\epsfig{file=Figs/dRl1j-BB.eps,height=5cm,clip=} & \quad &
\epsfig{file=Figs/dRl2j-BB.eps,height=5cm,clip=}
\end{tabular}
\caption{Distribution of several kinematical variables for $\mu^\pm \mu^\pm$
events from heavy neutrino production and dilepton events from $b \bar b$
production (see the text). }
\label{fig:bbsplit}
\end{center}
\end{figure}

Apart from these theoretical uncertainties there is another one due to the
limited statistics of the simulated samples. The $b \bar b$
cross section is 1.4 $\mu$b when both $b$ quarks are required to have
$p_T^b \geq 20$ GeV
at the generator level.
Fast simulation of 30 fb$^{-1}$ would take about 15000
days in a modern single-processor system,
making this computation difficult even in multi-processor grids.
(Full simulation would take about
$10^6$ years and, as emphasised above, in the real experiment this background
must be estimated from data, as it has been done by D0 \cite{D0}.)
Therefore, for our evaluations we have simulated samples of approximately 100,
35, 15 and 5 million events
for $b \bar b$, $b \bar b j$, $b\bar b 2j$ and $b \bar b 3j$,
respectively, corresponding to a luminosity $L = 0.075$ fb$^{-1}$ and the
cross sections given by ALPGEN. The size of the samples is reduced when
performing the MLM matching, which has efficiencies of 90.7\%, 41.8\%,
18.7\% and 12.7\%, respectively. The number
of events at pre-selection is calculated by rescaling the number
of events to 30 fb$^{-1}$. For example,
\begin{equation}
N(\mu^\pm \mu^\pm;\text{pre},30) \simeq N(\mu^\pm \mu^\pm;\text{pre},L) f_L \,,
\end{equation}
with $f_L = 400$. This rescaling introduces a large statistical uncertainty
and, moreover,  the estimation of the number of
events after selection cuts cannot be done in this way, since no $\mu^\pm
\mu^\pm$ events survive the cuts applied.
Instead, we make the reasonable assumption
that selection cuts, which are based on kinematical variables, have the same
effect on all $\ell \ell'$
events, where $\ell,\ell'=e,\mu$, not necessarily of the same charge.
Then, for $b \bar b nj$ backgrounds the number of $\mu^\pm \mu^\pm$ events after
selection cuts for 30 fb$^{-1}$ can be estimated from the samples with a smaller
luminosity $L$ as
\begin{equation}
N(\mu^\pm \mu^\pm;\text{sel},30) \simeq N(\ell \ell';\text{sel},L)
\left[ \frac{N(\mu^\pm \mu^\pm;\text{pre},L)}{N(\ell \ell';\text{pre},L)} f_L
\right] \,.
\label{ec:rescale}
\end{equation}
Since the total number of $\ell \ell'$ events is about $200$ times larger than
the number
of $\mu^\pm \mu^\pm$ events, the term in brackets in Eq.~(\ref{ec:rescale})
is of order two, and thus the simulated samples provide a statistically more
precise estimate
of the results for $\mu^\pm \mu^\pm$ final states. We have explicitly checked
whether the relevant kinematical distributions are similar or not for several
final states.
In particular, differences between electrons and muons might be expected due
to the different energy resolution and isolation criteria.
The most crucial variables for background suppresion are the transverse momenta
of the two leptons. They are presented in Fig.~\ref{fig:bbsplit}, together with
a ``signal'' sample
included for comparison. The heavy neutrino sample corresponds to more than
40000 events, while the dilepton samples from $b \bar b nj$ contain 37
$\mu^\pm \mu^\pm$, 382 $\mu^+ \mu^-$, 1497 $e^\pm e^\pm$ and 4676 $e^+ e^-$
events, respectively. The $\mu^+ \mu^-$, $e^\pm e^\pm$ and $e^+ e^-$ samples
have remarkably similar distributions, while $\mu^\pm \mu^\pm$ events
apparently concentrate at lower transverse momenta. This seems to be only a
statistical effect, given the smallness of the sample (only 37 events). This
belief is strengthened if one realises that $e^\pm e^\pm$ and $e^+ e^-$ events
have the same distributions (what suggests charge independence) and the same
happens for $e^+ e^-$ and $\mu^+ \mu^-$ (suggesting flavour independence).
Two further variables which might exhibit differences are the distance between
the leptons and the closest jet, also shown in Fig.~\ref{fig:bbsplit}. In this
case there seem to be small differences between the samples. However, these two
variables are not determinant in suppressing the background, as it can be
observed by comparing with the $N$ signal distribution, and any eventual
difference in kinematics will have little effect on our calculations.

Finally, it is worth remarking here
that raising the $p_T^b$ threshold at generator level, {\em e.g.} to 50 GeV,
leads to a dramatic
reduction of the $b \bar b nj$ cross sections,
making the simulation more manageable. However, this
also results in a gross underestimation of the $b \bar b nj$ background.
We have checked this by simulating two samples of 25 million $b \bar b$ events
with $p_T^b \geq 20$ and $p_T^b \geq 50$ GeV, respectively.
For pre-selection we just require two
isolated muons of either charge with $p_T^\mu \geq 10$ GeV.
For the $p_T^b \geq 20$ sample we obtain 364 $b \bar b \to \mu \mu$ events,
while for the $p_T^b \geq 50$ sample we only obtain 287 events.
Given the difference in cross sections (1430 nb for $p_T^b \geq 20$ GeV
and 58.8 nb for $p_T^b \geq 50$ GeV),
raising $p_T^b$ to 50 GeV at event
generation would underestimate this backgrounds by a factor of 30.
For $b \bar j$ we have checked that raising $p_T^b$ at event generation
to 50 GeV reduces the number of dimuon events by a factor of 35.
This seems to be related to the fact
that $b$ quarks with larger transverse momentum give more collimated decay
products, and thus the muons are less isolated. On the other hand, $b$ quarks
with too low transverse momentum cannot produce muons with
$p_T^\mu \geq 10$ GeV.
For this reason, we expect that our evaluation of $b \bar b nj$ provides a good
estimate of the actual background to be found at LHC.

The evaluation of
from $c \bar c nj$ proceeds in the same way. However, the number of dilepton
events is much smaller and no $\mumu$ events appear
in the samples simulated (containing about 145 million events after MLM
matching). In this case the
number of $\mumu jj$ events from $c \bar c nj$ production is estimated by
comparing with $\ee nj$ events and assuming the same ratio of events
$N(\mumu jj)/N(\ee jj)$ as for $b \bar b nj$ production.
The result is shown between parentheses in the tables.

\section{Heavy neutrino mass reconstruction}
\label{sec:b}

\begin{figure}[htb]
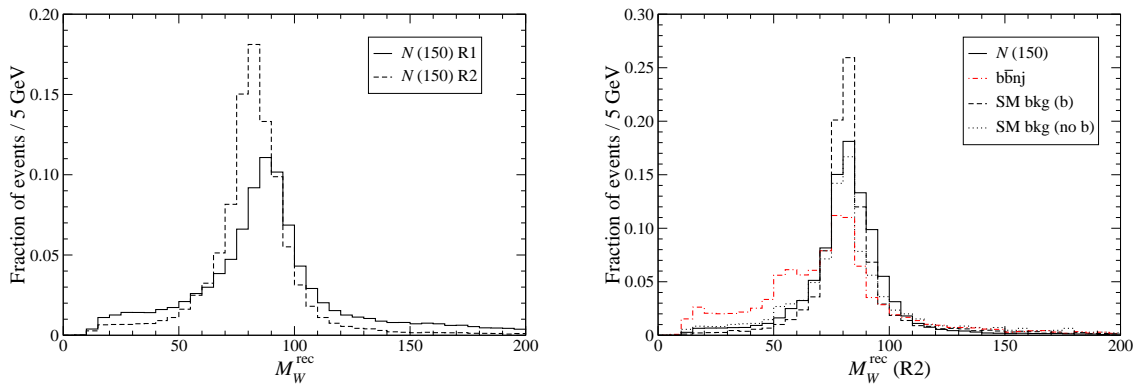

\begin{center}
\begin{tabular}{ccc}
\epsfig{file=Figs/MWcomp-N.eps,height=5cm,clip=} & \quad &
\epsfig{file=Figs/MWrec-bkg.eps,height=5cm,clip=}
\end{tabular}
\caption{Left: reconstructed $W$ mass for the heavy neutrino signal, using the
two methods (R1 and R2) explained in the text. Right: reconstructed $W$ mass
for the signal and SM backgrounds using the method R2.}
\label{fig:rec-comp}
\end{center}
\end{figure}

For heavy neutrinos $N$ heavier than the $W$ boson the decay
$N \to \mu W \to \mu q \bar q'$ takes place with $W$ on shell; thus, the
invariant mass of the two quarks is
$M_W$ up to finite width effects. In simulated signal events, however, several
extra jets often appear due to pile-up and initial/final state radiation, and
it is not straightforward to identify the two ones originating from the $W$
decay. We have tested two procedures:
\begin{enumerate}
\item To take, naively, the two jets with highest transverse momentum. This
method will be denoted as `R1'.
\item To try all possible pairings among the jets, choosing the pair giving an
invariant mass closest to $M_W$. This method is denoted as `R2'.
\end{enumerate}
The difference between the two choices is illustrated in Fig.~\ref{fig:rec-comp}
(left) for the case of the heavy neutrino signal.
The method R1 yields a moderate peak for the $W$ reconstructed mass
$M_W^\text{rec}$. When included in likelihood function (see
Fig.~\ref{fig:vars2}), this variable improves the signal significance about
2\%. (No improvement is found when performing a kinematical cut on
$M_W^\text{rec}$ in addition to the cut on likelihood.)
The second method R2 gives a considerably more peaked distribution for the
signal, at the expense of strongly biasing the background, as it is shown in
Fig.~\ref{fig:rec-comp} (right). Thus, with the second method the $W$ invariant
mass is not a useful variable for discriminating the signal against the
background.

\begin{figure}[htb]
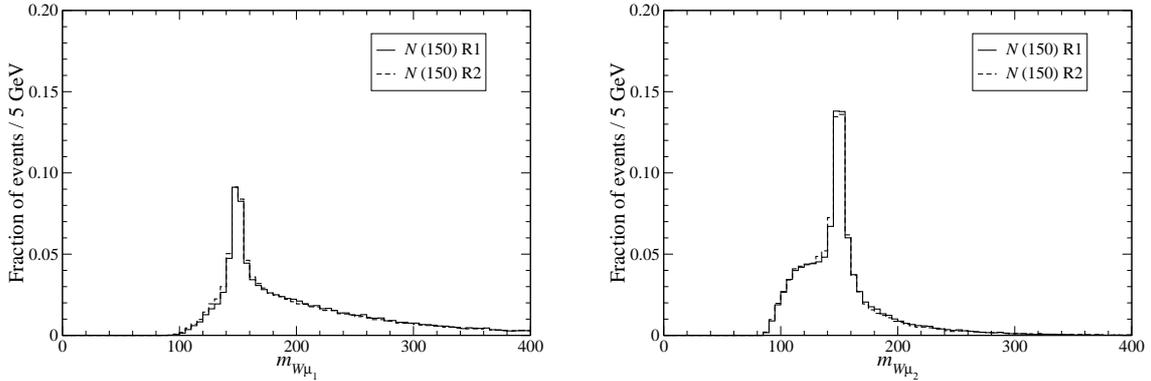

\begin{center}
\begin{tabular}{ccc}
\epsfig{file=Figs/mN1-comp.eps,height=5cm,clip=} & \quad &
\epsfig{file=Figs/mN2-comp.eps,height=5cm,clip=}
\end{tabular}
\caption{Invariant mass of the reconstructed $W$ boson (rescaled) and the muon
with highest ($\mu_1$) and lowest ($\mu_2$) transverse momentum, for the two
$W$ reconstruction choices explained in the text.}
\label{fig:rec-comp2}
\end{center}
\end{figure}

The heavy neutrino mass is obtained as the invariant mass of the jet pair
selected to reconstruct the $W$, plus one of the two muons.
In order to improve the reconstruction, the two jet
momenta are rescaled so that their invariant mass coincides with $M_W$. 
For both $W$ reconstruction methods the results are very similar, as it can
be observed in
Fig.~\ref{fig:rec-comp2}. The invariant mass of the $W$ and the muon with
smaller transverse momentum $m_{W\mu_2}$ is more concentrated around the true
$m_N$, and is taken as the heavy neutrino reconstructed mass in our analysis.
In case of discovery, this distribution might be used to determine $m_N$.

\begin{figure}[htb]
\begin{center}
\epsfig{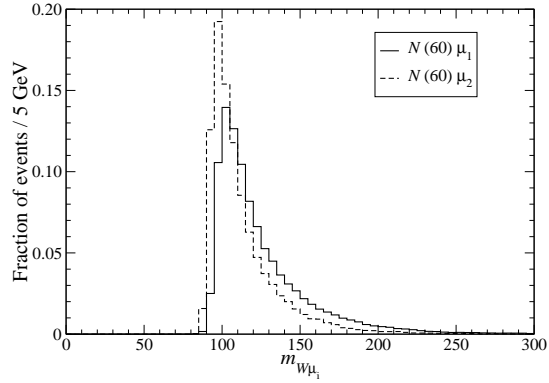} 
\caption{Invariant mass of the two jets with highest transverse momentum and 
each of the two muons, for a heavy neutrino signal with $m_N = 60$ GeV.}
\label{fig:mWl-N60}
\end{center}
\end{figure}

For $m_N < M_W$ it is very difficult to identify the two jets coming from
$N \to W^* \mu$, which have low transverse momenta, due to the appearance of
extra jets from pile-up. This fact is clearly seen
examining the invariant mass distribution of the two jets with highest $p_T$ 
and either of the two muons, in Fig.~\ref{fig:mWl-N60}.
In both cases the distribution peaks well above
$m_N = 60$ GeV, indicating that one or the two jets do not really originate
from the heavy neutrino decay. We have not found any improvement of the signal
significance considering these variables in the likelihood analysis.

\begin{figure}[htb]
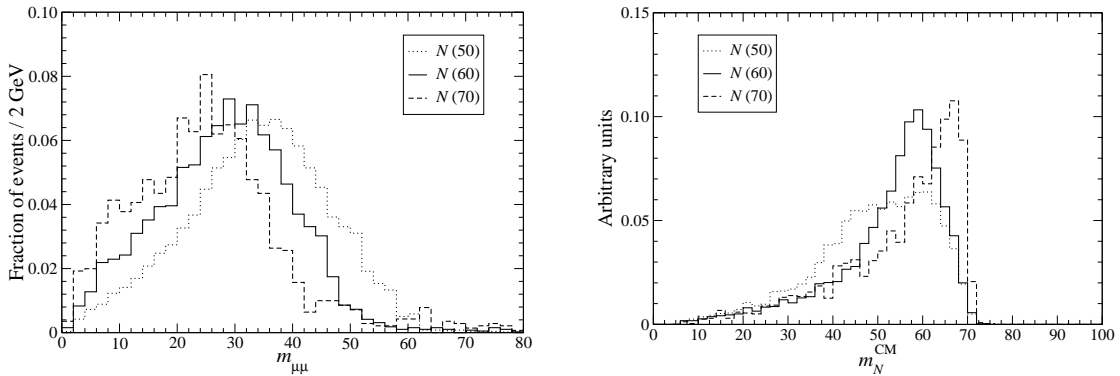

\begin{center}
\begin{tabular}{ccc}
\epsfig{file=./Figs/mll-mN.eps,height=4.9cm,clip=} & &
\epsfig{file=./Figs/mNcm.eps,height=4.9cm,clip=} \\
\end{tabular}
\caption{Left: invariant mass of the two muons, for three heavy neutrino masses.
Right: heavy neutrino mass reconstructed from the muon energy in CM frame.}
\label{fig:mNrec}
\end{center}
\end{figure}

In case of discovery, one possibility for the $N$ mass determination could be
to consider the
invariant mass of the two muons, which we present in Fig.~\ref{fig:mNrec} (left)
for heavy neutrino masses of 50, 60 and 70 GeV. This distribution seems to
peak around $m_N/2$. Other possibility is to exploit the fact that, since the
on-shell decay $W \to
\mu N$ is two-body, the energy of this muon in the centre of mass
(CM) system, $E_\mu^\text{CM}$, is fixed by $m_N$. Thus, we may determine the
heavy neutrino mass as
\begin{equation}
m_N^\text{CM} = \sqrt{M_W^2 - 2 M_W E_\mu^\text{CM}} \,.
\label{ec:mNrec}
\end{equation}
The $m_N$ reconstruction from the muon energy in the CM frame (defined as the
rest frame of the two muons and two jets with largest $p_T$) is shown in
Fig.~\ref{fig:mNrec} (right). For each event two values of $m_N^\text{CM}$ are
calculated, corresponding to the two possible muon choices,
and both are plotted. Imaginary values
are discarded.
These procedures for $m_N$ determination will be subject to
possibly large systematic uncertainties, but their evaluation is beyond the
scope of this work. (For example, the reconstruction from the muon energy in
the CM
frame is expected to have a systematic uncertainty from pile-up, which could be
decreased using the muon energy in the laboratory frame, but at the expense of
losing sensitivity to $m_N$.)
If heavy neutrinos were discovered, interesting information about CP violation,
relevant for leptogenesis, could also be inferred \cite{Bray:2007ru}.

\end{document}